\documentclass[superscriptaddress, amsmath,amssymb, reprint, aps, prfluids]{revtex4-2}

\usepackage{comment}
\usepackage[usenames,dvipsnames]{xcolor}
\usepackage[hidelinks]{hyperref}
\usepackage[normalem]{ulem}
\usepackage{bm}

\newcommand{\DPS}{\Delta P^\star}

\usepackage{tikz}
\usetikzlibrary{calc}
\definecolor{revisions}{RGB}{0, 0, 0}
\definecolor{revisions2}{RGB}{0, 0, 0}
\definecolor{revisions3}{RGB}{0, 0, 0}

\definecolor{referee1}{RGB}{0, 0, 0}
\definecolor{referee2}{RGB}{0, 0, 0}

\begin{document}

\title{Hydrodynamic resistance of a yeast clog}%

\author{T. Desclaux}
\affiliation{Institut de Mécanique des Fluides de Toulouse, Université de Toulouse, CNRS, Toulouse, France. }
\affiliation{LAAS-CNRS, Université de Toulouse, CNRS, Toulouse, France.}
\author{L. Santana}
\author{I. Verdeille}
\author{P. Duru}
\affiliation{Institut de Mécanique des Fluides de Toulouse, Université de Toulouse, CNRS, Toulouse, France. }
\author{P. Joseph}
\author{M. Delarue}
\affiliation{LAAS-CNRS, Université de Toulouse, CNRS, Toulouse, France.}
\author{O. Liot}
\email{Corresponding author: olivier.liot@imft.fr}
\affiliation{Institut de Mécanique des Fluides de Toulouse, Université de Toulouse, CNRS, Toulouse, France. }
\affiliation{LAAS-CNRS, Université de Toulouse, CNRS, Toulouse, France.}

\date{\today}%

\begin{abstract}
Bioclogging, the clogging of pores with living particles, is a complex process that involves various coupled mechanisms such as hydrodynamics and particle properties. The lack of sensitive methods to simultaneously measure the hydraulic resistance of a clog and the position of particles at high enough resolutions limits our understanding of this tight fluid-structure interplay. In this article, we explored bioclogging at the microscale level, where flow rates are typically extremely low ($<100$ nL.min$^{-1}$). We developed a highly sensitive method to precisely measure small flow rates ($<6.7 \%$ error) with a low response time ($<0.2$ s) while retaining the capability to image the structure of a fluorescently-labeled yeast {clog} at high resolution. This method employed a microfluidic device with two identical channels: a first one for a yeast suspension and a second one for a colored culture medium. These channels merged into a single wide outlet channel, where the interface of the colored medium and the medium coming from the yeast clog could be monitored. As a yeast clog formed in the first channel, the displacement of the interface between the two media was imaged and compared to a pre-calibrated image database, quantifying the flow through the clog. We carried out two kinds of experiments to explore the fluid-clog interaction: filtration under constant operating pressure and filtration under oscillating operating pressure (backflush cycles). We found that in both scenarios, the resistance increased with clog length. On the one hand, experiments at constant imposed pressure showed that the clog's permeability decreased with increased operating pressure, with no detectable changes in cell density as assessed through fluorescence imaging. In contrast, backflush cycles resulted in an approximately four times higher permeability, associated with a significant non-monotonic decrease in cell density with the operating pressure. Leveraging on our unique dual measurement of fluid flow and clog microstructure, our study offered enhanced insights into the intricate relationships between clog microstructure, permeability, and construction history. A better understanding of the fluid-structure interplay allowed us to develop a novel physical modeling of the flow in a soft and confined porous medium that challenged the empirical power-law description used in bioclogging theory by accurately replicating the measured permeability-pressure variations.  

\end{abstract}

\maketitle

\section{Introduction}
%%%%%%%%%%%%%%%%%%%%%%%%%%%%%%%%%%%%%%%%%%%%%%%%%
%%%%%%%%%%%%%%%%% INTRODUCTION %%%%%%%%%%%%%%%%%%
%%%%%%%%%%%%%%%%%%%%%%%%%%%%%%%%%%%%%%%%%%%%%%%%%

Liquid chromatography, separation of cells from fermentation broths in many industrial processes   \citep{meireles_filtration_2003, foley_review_2006}, or quantification of the abundance of oceanic plankton  \citep{hunt_continuous_2006} all rely on the separation of solid biological particles suspended in a liquid. A filtration process classically accomplishes this separation: the suspension flows through a porous membrane, which retains the particles and lets the fluid pass through. However, this process has a major disadvantage: the membranes can foul due to pore clogging, and the amount of accumulated material can become so large that the liquid flow rate becomes very low, almost null. According to the situation, this clogging can be desired, for example, to isolate specific objects such as cancer cells with diagnostic purpose \citep{tang_microfluidic_2015, pang_deformability_2015}, or to be avoided, for example, to treat wastewater \citep{in-soung_membrane_2002,thuy_impact_2022}.

Understanding the processes involved in pore clogging is complex, as it is governed by several coupled mechanisms: hydrodynamics, Brownian diffusion, surface interactions, the mechanical properties of the (living) particles, and aging. It is only recently that the general framework of clogging at the microscale with inert and rigid particles has begun to be well described  \citep{dressaire_clogging_2017,bouhid_de_aguiar_microfluidics_2020}, thanks to microfluidic studies reporting processes occurring at the scale of a single pore  \citep{wyss_mechanism_2006,bacchin_colloidal_2011,dersoir_clogging_2015,sauret_growth_2018,mokrane_microstructure_2020,vani_influence_2022}, or of a single particle  \citep{lin_particle_2009, duru_three-step_2015,cejas_particle_2017,cejas_universal_2018,dersoir_clogging_2017,  delouche_flow_2022}. These different studies detail the mechanism of capture of particles by pore surface and highlight how the competition between Brownian motion, drag forces and physicochemical surface interactions affect particle capture, clog initiation, and growth dynamics.

Recently, several studies documented the use of microfluidic devices to analyze clogging with soft materials such as oil droplets \citep{hong_clogging_2017} or hydrogels \citep{linkhorst_direct_2019,luken_particle_2021-1}. Nevertheless, the study of bioclogging - clogging by biological objects - remains a current research challenge. Indeed, living cells are deformable, polydisperse, endowed with specific adhesion mechanisms \cite{el-kirat-chatel_forces_2015}, sensitive to mechanical forcing \citep{alric_macromolecular_2022}, and are alive (i.e., consume nutrients, can be mobile, divide, differentiate, die). 
In particular, predicting the hydrodynamic resistance of a living clog is notoriously difficult. Theoretical predictions are still out of reach, while experimental values obtained with clogs made of living cells may vary by several orders of magnitude from one study to the other \citep{foley_review_2006}. In particular, the specific resistance - defined as the hydrodynamic resistance divided by the mass of the cake per unit membrane area – depends on the operating pressure pushing the suspension through the membrane \citep{foley_review_2006}. For example, the specific resistance of a filtration cake made of \textit{Saccharomyces cerevisiae} ranges from $0.9\times10^{11}$\,m.kg$^{-1}$ at 4 kPa to $3\times10^{11}$\,m.kg$^{-1}$ at 12.8 kPa in Valencia et al.  \citep{valencia_direct_2022}. This variation of hydrodynamic resistance with hydrodynamic pressure is interpreted to be a consequence of the compressibility of biological clogs, a combined effect of particle rearrangements, and deformation \citep{foley_review_2006}. However, no scientific consensus exists on this subject. The study by Ben Hassan et al., with observations at the microscale, \citep{ben_hassan_study_2014} reported no compression of the clog: the clog height was constant, whatever the hydrodynamic pressure imposed. Another study \citep{meireles_origin_2004}, based on numerical simulations, reported that taking into account the clog compressibility is not sufficient to explain the high hydrodynamic resistance observed with living clogs, but that the interactions between the first cells and the membrane contribute to a large part of the hydrodynamic resistance. Finally, a very recent study \citep{valencia_direct_2022} showed that yeast clogs are indeed compressible and that this compressibility is important to explain the hydrodynamic resistance of the clog.  
These studies highlight the importance of the mechanical and biological interactions at play within a clog of deformable yeast cells.  This is also evidenced when a backflush of the clog is performed. During a backflush, the direction of the flux throughout the membrane is reversed to remove the clogged particles and to recover a high enough permeate flux \citep{park_effects_1997,matsumoto_separation_1987,smith_new_2006, smith_design_2005,mugnier_optimisation_2000,park_effects_1997}.  At the microscale, Lohaus et al. \citep{lohaus_what_2020} recently provided a precise understanding of the mechanisms that drive backflushes. They highlighted that, for rigid colloids, three different phenomena occur during a backflush: (i) resuspension of individual particles, (ii) resuspension and re-orientation of clusters of particles, and (iii) fragmentation of clusters of particles. Another study reported the occurrence of both the resuspension of clusters and further fragmentation of clusters with soft microgels \citep{luken_unravelling_2020}. Finally, the resuspension of particle clusters was confirmed, as these clusters may accelerate the clogging of parallel channels \citep{dincau_clog_2022}. Backflushes can then be seen as a way to test the mechanical/cohesive properties of a clog and also to modify the microstructure of the suspension to be filtered (aggregates vs. isolated particles) when the direction of the flow is restored to its original direction.

In this paper, we present observations of bioclogging at the microscale, to better understand the variability of the hydrodynamic resistance with the operating pressure in relation to direct visualizations of the clog. A microfluidic device is used to mimic a single-pore filtration membrane.  Yeast \textit{S. cerevisiae} is chosen as a model biological object whose biology \citep{alric_macromolecular_2022} and mechanics \citep{vella_indentation_2012} are well described, while the experimental protocol is designed to avoid any osmotic shock for the yeasts susceptible to alter the cells mechanical behavior during the experiment. As a pre-requisite, we have developed and validated a method to measure low microfluidic flow rates, inspired by Choi et al. \cite{choi_microfluidic_2010} and relying on a specific design of the microfluidic chip. Indeed, because of the micrometric scales involved and the high hydrodynamic resistance of clogs, the flow rate inside microfluidic devices rapidly reaches very low values ($< 100$ nL.min$^{-1}$), well below the range of available commercial flowmeters.  We present experiments focussing on the formation and properties of yeast clogs, conducted at fixed pressure, complemented with backflush experiments designed to investigate the effect of the yeast suspension microstructure on the clog properties.  
The paper is organized as follows. The experimental device and methods are presented in the section \ref{sec:MatEtMet} with a highlight on our homemade flow rate measurement method. Section \ref{section:Results} presents results of hydrodynamic resistance and permeability during the yeast clog construction at fixed pressure and during backflush cycles. Section \ref{sec:Model} details a model to represent the variability of the permeability as the operating pressure increases. Section \ref{sec:Discussion} presents a discussion of the main results of this article, and section \ref{sec:Conclusion} the conclusions and some perspectives opened by the present work.

\section{Material and methods}
\label{sec:MatEtMet}
%%%%%%%%%%%%%%%%%%%%%%%%%%%%%%%%%%%%%%%%%%%%%%%%%
%%%%%%%%%%%%%%%%% M&M %%%%%%%%%%%%%%%%%%%%%%%%%%%
%%%%%%%%%%%%%%%%%%%%%%%%%%%%%%%%%%%%%%%%%%%%%%%%%
\subsection{Experimental methods}
\subsubsection{Yeast suspension and colored solution}
\textit{Saccharomyces cerevisiae} yeasts are used as a biological model. These single-celled organisms are ellipsoidal and have an average radius of around $ 2.65\,\mu$m. The {``HTB2-mCherry''} strain used is genetically modified to display a fluorescent nucleus. The yeast culture is done at room temperature, and the cell population doubles every 2h30 to 2h45. 

A liquid culture solution is prepared so that the mass density of the solution is equal to the mass density of yeast cells, and so that the osmolarity of the medium is equal to the osmolarity of classical culture medium:
$ 13\% $ of distilled water (percentage given by volume), $ 39\% $ iodixanol solution concentrated at $ 60\%$ (w/v)  (\textit{Optiprep}), $ 48\% $ solution of Synthetic Complete Dextrose Medium (SCD) concentrated at 2 $\% $ in dextrose (D-glucose). The dynamic viscosity of this culture media is $\eta =2\times10^{-3}$\,Pa.s. A colored solution is prepared with the same protocol as the isodense culture medium but replacing pure distilled water with an already colored solution (\textit{AQUAcouleur\texttrademark, Ocean color}).

The cells are grown in a Petri dish on a 0.5\,cm thick layer of agarose (agar-agar mixed with Synthetic Complete Dextrose medium). The day before experiments, the cells are resuspended: they are extracted from the Petri dish and mixed with the liquid culture medium, and a series of dilutions is carried out. The cells are left in culture in the laboratory overnight. The next day, the tube containing a concentration typically in the range $ [5\times10^5 - 10^6]\,$cells.mL$^{-1}$ is used. The yeast concentration in the suspension is checked using a spectrophotometer. The device used measures the absorbance of the solution at a wavelength of 600\,nm. The cell concentration is calculated according to \citep{fukuda_apparent_2023}: $ C_{cell} \approx 10^7 \times OD_{600}$ cells.mL$^{-1}$, where $ C_{cell} $ is the number of cells per mL in the suspension and $OD_{600}$ is the absorbance value.

In our experiments, the proliferation of yeasts is stopped by using an antibiotic (cycloheximide - CHX). The antibiotic is added to the yeast suspension at $ 10\,\mu$M one hour before flowing the suspension through the microfluidic device. It blocks the synthesis of proteins \cite{delarue_mtorc1_2018}, and thus cell proliferation, {\color{referee1} without any expected changes in the cell adhesion and mechanical properties}. However, cells are still alive, as when the antibiotic is rinsed, cells proliferate again. 

{It is important to note that great care was taken in avoiding any osmotic shock that could be experienced by the cells and which would modify their mechanical properties (e.g. elasticity \citep{vella_indentation_2012}). Indeed, living cells maintain turgor thanks to osmotic differences with the surrounding medium:  these differences in osmolarity create a difference in chemical potential that draws water inside the cells. They are then swollen, and the excess interior pressure is called the turgor pressure. If the osmolite concentration outside the cell increases, then the difference in chemical potential decreases, and so does the turgor pressure inside the cell: the cell deflates.  }

\subsubsection{Microfluidic chips and flow control}
The micromodel is shown schematically in Fig. \ref {fgr:bouchon}. The chip is etched in silicon using standard photolithography and plasma etching in a clean room. \textcolor{referee1}{Silicon/glass chips are preferred to more standard Polydimethylsiloxane (PDMS) ones to avoid possible material deformation during the experiments.} The etching depth is precisely characterized by a mechanical profilometer with about 1\% uncertainty. Two groups of chips are used in this study. The first group is etched at depth of $H \in [6.05, 6.15]\,\mu$m, while the other has a depth of $H \in [6.3, 6.4]\,\mu$m. Two identical side-by-side inlet microchannels of width $W=140\,\mu$m and 6.6\,mm long are designed on each chip. At the end of each microchannel, a constriction of $w=6\,\mu$m wide and $L=6\,\mu$m long acts as a pore to allow yeast clogging. After the constrictions, the microchannels are merged in a unique 315\,$\mu$m wide and  3.35\,mm long outlet microchannel. The chip is bonded using a 170-$\mu$m thick borosilicate layer by anodic bonding. 

{Each inlet and outlet of the microfluidic chip is connected directly to a reservoir by microfluidic tubing of 0.5 mm inside diameter and 40 cm in length. Pressure controllers (\textit{Fluigent Flow EZ\texttrademark}) then apply the required hydrodynamic pressure into these reservoirs.}

\subsubsection{Clog observation}

\begin{figure}
\centering
  \includegraphics[width=\linewidth]{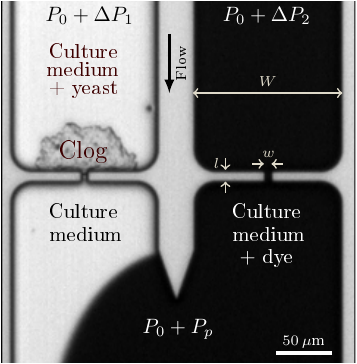}
  \caption{Micrography of the microsystem ($5 \times$  magnification). The yeast suspension is pushed in the left-hand microchannel at pressure  $P_0+\Delta P_1$, with $P_0$ the atmospheric pressure. The constriction retains the yeast particles, so a clog is constructed on the left-hand microchannel. Colored culture medium flows in the right-hand microchannel at pressure $P_0 + \Delta P_2$. An interface between clear and colored liquids appears downstream of the merging region, where the pressure equals $P_0 + P_p$. The location of this interface is used to measure the flow rate ratio between the two microchannels \cite{coyte_microbial_2017}. Each microchannel has a width $W = 140\, \mu$m. The constriction has a width $w = 6\, \mu$m and a length $L=10\, \mu$m, and the merging region has a width equal to $315\, \mu$m.
}
  \label{fgr:bouchon}
\end{figure}

\textcolor{referee1}{The confinement ratio between channel depth and typical yeast diameter is slightly larger than one. Consequently, the clog is a quasi-2D yeast assembly.}
The clog formation and geometrical characteristics are captured using a \emph{Zeiss A.1} microscope. It is equipped with a monochromatic light source (\emph{Spectra X Lumencor}). The microsystem is illuminated using a red wavelength ($640\pm30\,$nm) known to have low photo-toxicity \cite{schmidt_preventing_2020}. Yeasts have a very similar refractive index as the culture medium we use. To enhance the contrast of the clogs, Hoffman contrast modulation  \citep {hoffman_modulation_1975} is used with a \emph{Zeiss EC Plan Neofluar} objective ($20\times$ magnification). An example of an acquired image is shown in Fig. \ref{fgr:bouchon}, where a clog is visible in the left-hand microchannel - in this example, a 5$\times$ magnification was used.  

A \emph{LaVision’s Imager sCMOS CLHS} camera is used for image acquisition, with a detector size of 2560$\times$2160\,px$^2$ leading to a 0.33 microns pixel size (at 20 $\times$ magnification). For clog growth, we use a framerate of 1 frame per second. Exposure time is set to $2000\,\mu$s. Note that illumination is synchronized with camera acquisition to reduce the light dose experienced by the yeasts and photo-toxicity.

\subsubsection{Image analysis}
\label{subsec:ImageAnalysis}
A dedicated algorithm has been set to measure the clog length. This algorithm consists of edge detection followed by several morphological operations. The shape of the clog interface is quasi-semicircular at the beginning of the experiments (see Fig. \ref{fgr:bouchon}). Once the clog radius has reached the lateral limit of the channels, it adopts a quasi-rectangular shape (see Fig. \ref{res:ConstructionIllustration}). The clog length ($L_c$) is defined as the minimal distance between the pore and the top border of the clog (see a good example in Fig. \ref{res:ConstructionIllustration}\,F). In the worst-case scenario, this algorithm is estimated to be precise by one or two cell diameters (see inset in Fig \ref{res:ConstructionIllustration}\,A). Therefore, in the following, the uncertainty in the clog length is estimated to be  10 $\mu$m.

\begin{figure}
\centering
  \includegraphics[width=\linewidth]{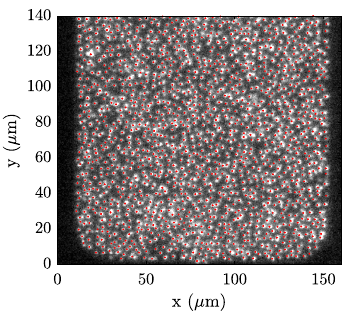}
  \caption{ Raw fluorescence pictures and detected particles. The raw fluorescence image is displayed in black and white. A red dot represents the position of a detected cell. For visualization, only the first 140 $\mu$m of the clog are represented. This clog has been constructed at constant pressure, with an operating pressure of 40 kPa.}
  \label{res:MICRO-Illustration}
\end{figure}

Besides, because the HT2B-mCherry yeast cells produce fluorescent histones, it is possible to image their nuclei by fluorescence microscopy (one example image is presented in Fig. \ref{res:MICRO-Illustration}). \textcolor{referee1}{As the clog is quasi 2D, fluorescence microscopy can image all the yeasts in the clog depth.} A dedicated algorithm has been adapted from a previous study \citep{mokrane_microstructure_2020} to automatically detect the position of these fluorescent nuclei that appear as white dots. The result is a list of positions, shown as red dots in Fig. \ref{res:MICRO-Illustration}. We finally compute the clog area by computing the polygon area that includes all these points (plus an external layer of $2.5\,\mu$m of width). The cell density ($C_c$) is then defined as the number of cells in a clog divided by its area times the height of the device. We estimate the uncertainty of this measurement to be 5\% (see Supplementary Information 1).

\subsection{On-chip flow rate measurement}
\label{section:OnChipFlowRate}

\begin{figure*}
\centering
  \includegraphics[width=\linewidth]{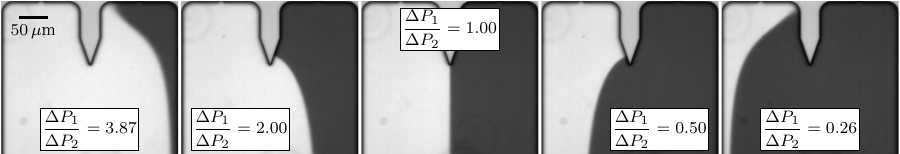}
  \caption{Examples of pictures extracted from the data bank. The pressure drop ratio is indicated in each picture. The corresponding $Q_1/Q_2$ ratios are, from left to right: 43.05, 3.35, 1.0, 0.30, 0.025.}
  \label{fgr:database}
\end{figure*}

\subsubsection{Motivations} 
{
Numerous methods have been developed to measure flow rates in micro- and nanosystems, see e.g. the recent and exhaustive review by Cavaniol et al. \citep{cavaniol_flowmetering_2022}. A distinction is usually made between \emph{active} and \emph{passive} methods. The first type of method is based on adding external energy to the system to retrieve information about the flow rate. Most commercial flow rate sensors are based on an active calorimetric detection \citep{ejeian_design_2019,cavaniol_flowmetering_2022}. They can measure small flow rates (under 100\,nL.min$^{-1}$) but with a limited measurement range and {\color{revisions3}more than 10 $\%$ uncertainty} for the lowest flow rates. Numerous passive methods, including seeding of particles, are commonly used in microfluidics. Nevertheless, this is not always adapted to applications that cannot accept contaminations with foreign objects. This is especially the case for the clogging process, where adding probe tracers could affect the clog's structure and the clogging dynamics. Furthermore, some methods based on particle seeding may need long acquisitions to get enough statistics to estimate the flow rate \citep{ranchon_metrology_2015}. Consequently, recent studies about clogging have used either commercial flow rate sensors when the flow rates are large enough \citep{dincau_clog_2022} or have measured directly the velocity of the particles flowing towards the clog \citep{delouche_flow_2022}. }

{In 2010, Choi et al. \citep{choi_microfluidic_2010} presented a microfluidic design allowing to compare, through direct visualizations, the hydrodynamic resistance of a test channel with the known resistance of a reference channel. The design is such that the two channels merge downstream the region of interest (e.g. a constriction in the test channel, where flow-driven particles accumulate), and is similar to the one presented in Fig. 1. The fluid in the reference channel is colored so that the streamline separating the flow coming from each channel is easily observable. The position of this streamline, relative to the side walls and far enough from the merging region, can be calibrated against the ratio of the flow rates in each channel. A similar method, with further theoretical developments, was used by Coyte et al. \citep{coyte_microbial_2017} to study microbial competition in two connected microchannels. }

{We have elaborated from these works to fit the specificities of the clogging process: very important flow rate decline and the need for a large field of view to track the clog growth. In particular, due to molecular diffusion of the coloring agent (see Fig. \ref{fgr:bouchon}), the location of the interface between the colored and no-colored fluids can lack precision, and when the interface is close to the wall, the difficulty is enhanced \cite{coyte_microbial_2017}. Moreover, in our configuration, having the whole clog in the field of view is compulsory. It may reach several hundreds of micrometers and even millimeters in length. In this case, it is impossible to have both the clog and the region where the interface is parallel to the microchannel walls on the same picture. We propose below an adaptation of the interface detection method based on image recognition, using the interface shape in the merging zone instead of its position relative to the walls.}

\subsubsection{Method calibration}
This method needs precise calibration to relate the shape of the interface to a flow rate ratio. A data bank of pictures is generated. The left-hand microchannel is filled with a transparent culture medium without yeast, and the right-hand one is filled with a colored culture medium, so the viscosities are identical. The pressure drops $\Delta P_1$ and $\Delta P_2$ applied in the left-hand and the right-hand microchannels, respectively, are varied to have a wide range of pressure ratio. The data bank comprises 191 pictures with a range of pressure drop ratios in [$0.25-3.74$]. This range cannot be extended much because, out of it, one observes a bypass flow from one of the two inlet microchannels to the second one, which is irrelevant for the calibration and clogging experiments. Fig. \ref{fgr:database} shows five pictures extracted from the data bank with respective pressure drop ratios. One database was created with an objective magnification of $20 \times$, as this magnification is used in this article. But note that, if needed, this database can be rescaled to treat images taken at other magnifications.

\begin{figure}
    \centering 
    \includegraphics[width=0.5\linewidth]{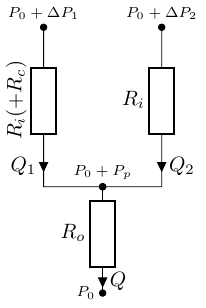}
    \caption{Equivalent electrical circuit of the microfluidic system in absence of clog. $Q_1$ and $Q_2$ are the flow rate in the left-hand and right-hand microchannels, respectively. $Q$ is the flow rate in the merged microchannel. $R_i$ is the hydrodynamic resistance of an inlet microchannel plus a constriction and $R_o$ is the hydrodynamic resistance of the unique outlet microchannel. $P_p$ is the pressure at the merging point and $P=0$ is the atmospheric pressure.}
  \label{fgr:circuit_vide}
\end{figure}

Since we know the geometrical characteristics of the microfluidic system, we can deduce the hydrodynamic resistance of each segment of the microchannels. We define $Q_1$ and $Q_2$ as the flow rate in the left-hand and right-hand microchannels, respectively. $Q$ is the flow rate in the merged microchannel. $R_i$ is the hydrodynamic resistance of an inlet microchannel including the constriction, and $R_o$ is the hydrodynamic resistance of the merged microchannel. The resistance of the microfluidic tubing is negligible compared to the device's resistance. $P_p$ is the pressure at the merging point. The equivalent electrical circuit is displayed in Fig. \ref{fgr:circuit_vide}, and the hydrodynamic resistance values are detailed in Table \ref{tbl:Resistances}. Using the electrical-hydrodynamic analogy, we can compute the flow rate ratio $Q_1/Q_2$ as a function of $\Delta P_1/\Delta P_2$.

\begin{table}
\centering
	\begin{tabular}{c|c|c}
	 & $R_i$ [Pa.s.m$^{-3}$] & $R_o$ [Pa.s.m$^{-3}$]\\
	 \hline
	 $H=6.1\,\mu$m & $5.55\times10^{15}$ & $1.72\times10^{15}$ \\
	 \hline
	 $H=6.35\,\mu$m & $4.94\times10^{15}$ & $1.52\times10^{15}$ 
	\end{tabular}
\caption{Table of values of the hydrodynamic resistances of each segment of the microfluidic system, for the two etching depths used in this study.}
\label{tbl:Resistances}
\end{table}
Using Kirchhoff's laws we show that:

\begin{equation}
    \dfrac{Q_1}{Q_2}=\dfrac{(\gamma+1)\dfrac{\Delta P_1}{\Delta P_2}-1}{(\gamma +1)-\dfrac{\Delta P_1}{\Delta P_2}},
    \label{eq:vide}
\end{equation}

\noindent with $\gamma=\dfrac{R_i}{R_o}$. Finally, the data bank corresponds to flow rate ratios in the range $[0.018 - 27.8]$.

\subsubsection{Image recognition}

To extract the flow rate ratio from an image taken during a clogging experiment, we first need to pre-process it. The experimental picture and the image data bank are cropped to have the same dimensions in pixels. A standard contour detection algorithm based on Radon transform is used to accurately detect the microchannels' walls.

\begin{figure}
\centering
  \includegraphics[width=\linewidth]{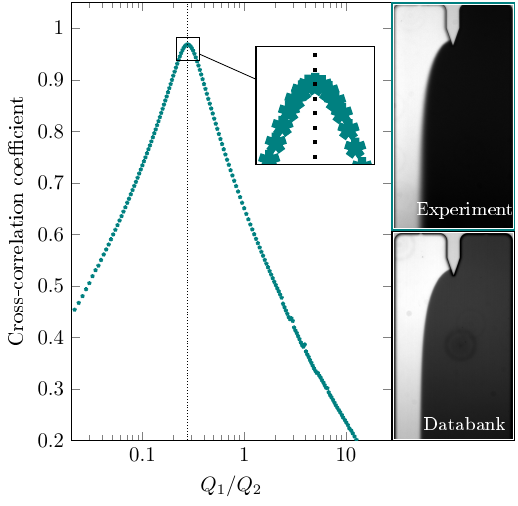}
  \caption{(left) Cross-correlation coefficient versus flow rate ratio of the pictures from the data bank. (top-right) Experimental picture. (bottom-right) Picture from the data bank corresponding to the cross-correlation {peak}.}
  \label{fgr:correlation}
\end{figure}

A standard two-dimensional cross-correlation algorithm is applied between the experimental picture and each data bank picture. This leads to a correlation coefficient associated with each flow rate ratio of the database. The cross-correlation coefficient presents a maximum for a given flow rate ratio, which we consider as the corresponding flow rate ratio. Fig. \ref{fgr:correlation}\,(left) provides an example of cross-correlation coefficients versus flow rate ratio. Fig. \ref{fgr:correlation}\,(top-right) shows the experimental picture, and Fig. \ref{fgr:correlation}\,(bottom-right) shows the image from the data bank corresponding to the maximal cross-correlation coefficient. 

\subsubsection{Clog resistance, method validation, and sensitivity}
\label{subsec:Const-Sensitivity}
In the presence of a yeast clog at the left-hand constriction, the clog's resistance $R_c$ is added to $R_i$, on the left-hand branch of Fig. \ref{fgr:circuit_vide}, and we deduce the following expressions for the clog's hydrodynamic resistance $R_c$, pressure at the merging point $P_p$, and flow rate through the clogged channel $Q_1$: 

\begin{equation}
R_c=(R_i+R_o)\left[\dfrac{\Delta P_1}{\Delta P_2}\dfrac{Q_2}{Q_1}-1\right]+R_o\left[\dfrac{\Delta P_1  }{\Delta P_2}-\dfrac{Q_2}{Q_1}\right],
    \label{eq:rhydr}
\end{equation}
\begin{equation}
P_p = \dfrac{ \Delta P_1 \dfrac{R_o }{R_i+R_c} + \Delta P_2 \dfrac{R_o}{R_i}}{1+\dfrac{R_o}{R_i+R_c}+\dfrac{R_o}{R_i} },
\end{equation}
\begin{equation}
Q_1 = \dfrac{P_1-P_p}{R_i+R_c}.
\end{equation}

Here, one can note that, when $R_c$ increases, it may become very large compared to $R_o$ and $R_i$. Then:
\begin{equation}
 P_p (R_c \gg R_o, R_i) \approx \Delta P_2 \dfrac{R_o}{R_i+R_o} \equiv P_p^\infty .
\end{equation}
{where $P_p^\infty$ represents the limit value of $P_p$, for large values of $R_c$. In that case, the difference of hydrodynamic pressure experienced by the clog is $\Delta P^\star = \Delta P_1 - P_p^\infty$.}

The method was tested and validated by randomly picking one picture from the data bank and applying the algorithm to this picture. By doing that, the expected result is known, and sensitivity tests can be done{, as detailed in Supplementary Information 1. Different tests have been performed to estimate the sensitivity of the method. The influence of blurring (corresponding to a loss of focus, for example), added noise (corresponding to noise induced by the camera, or defaults on the picture), interface zone length used for image recognition (with a centered or shifted interface in the microchannel) have no significant effect on the picture correspondence.}

{The uncertainty magnitudes associated with each of the variables introduced in this article are also quantified, see  Table \ref{tbl:Uncertainties}.  A detailed discussion on that point is given in Supplementary Information 1. 
}

\begin{table}
\centering
  \begin{tabular}{c|c|c}
	  Variable & Uncertainty sources & Magnitude \\
   \hline
   $L_c$ & Detection   &   10 $\mu$m \\
   \hline
   $C_c$ & Detection &  $5 \%$ \\
	\hline
	  $\Delta P_1, \Delta P_2,$ & \\
    $\Delta P^\star$, $P_p^\infty,$ & Pressure controllers & $<0.5\%$ \\
	\hline
	  $H$ & Fabrication & 1\% \\
	\hline
	  $\dfrac{Q_1}{Q_2}$ & Database discretisation & 6\% \\
	\hline
	  $Q_1, R_c$ & 3\% due to uncertainty on H & 6.7\% \\ 
	  & + 6\% due to database discretization &  \\
	\hline
	  k & 3\% due to uncertainty on H  & $\gtrsim3 \%$ \\
	  & + experiment-dependent uncertainty \\ 
	  & due to database discretization
   \end{tabular}
\caption{Uncertainty sources and magnitudes associated with each variable introduced in this article. {For more details, see Supplementary Information 1.} The definition of $k$ is introduced in section \ref{subsec:Permeability}.}
\label{tbl:Uncertainties} 
\end{table}

\begin{figure*}[t]
\centering
  \includegraphics[width=\linewidth]{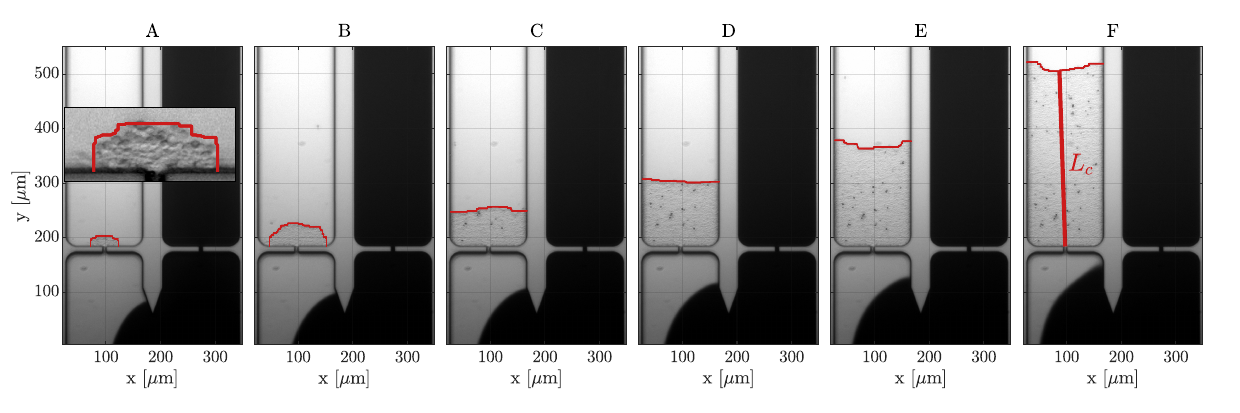} \\  
  \includegraphics[width=0.32\linewidth]{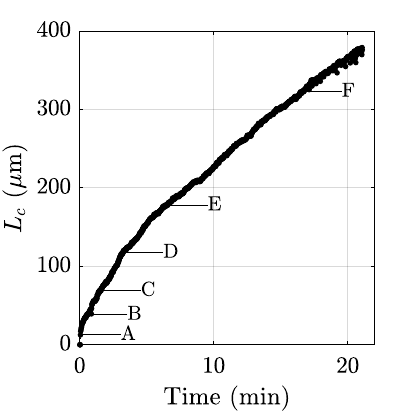}
  \includegraphics[width=0.32\linewidth]{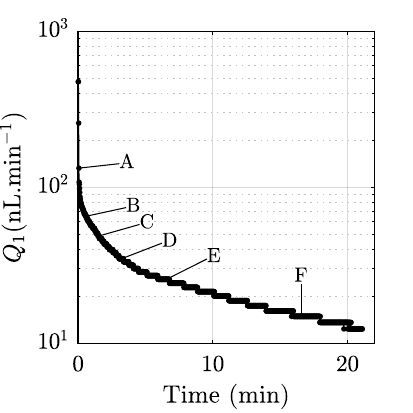}
  \includegraphics[width=0.32\linewidth]{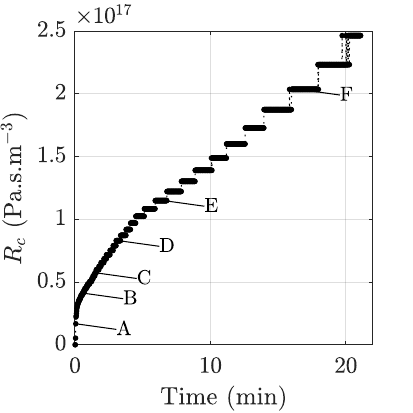} \\
  \caption{Top row: Simultaneous observation of the clog growth and of the flow rate reduction. From A to F: different images taken during a typical experiment. In the top part of each picture, the clog area is computed from clog segmentation (red lines). The inset of picture A represents a zoom on the clog. The clog length \textcolor{referee2}{(defined as the minimal distance between the pore and the top border of the clog)} of image F is displayed with a dark red line. Bottom row: clog length ($L_c$), flow rate ($Q_1$), and clog hydrodynamic resistance ($R_c$) evolution as a function of time after the beginning of clogging. Experience realized with $\Delta P_1 = \Delta P_2 = $ 70 kPa.}
  \label{res:ConstructionIllustration}
\end{figure*}

\begin{figure*}[t]
\centering
  \includegraphics[width=\linewidth]{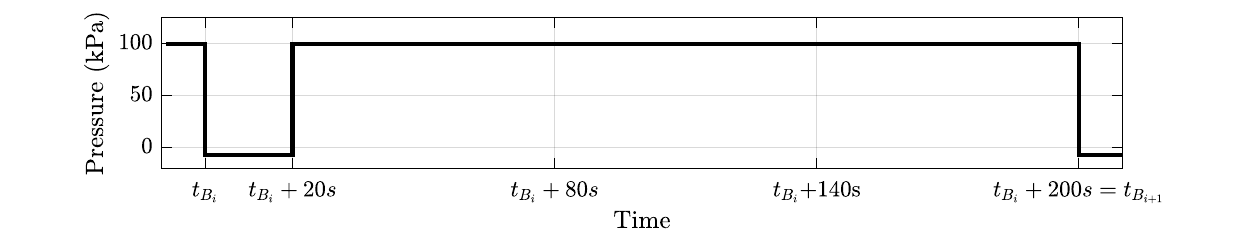} \\
  \includegraphics[width=\linewidth]{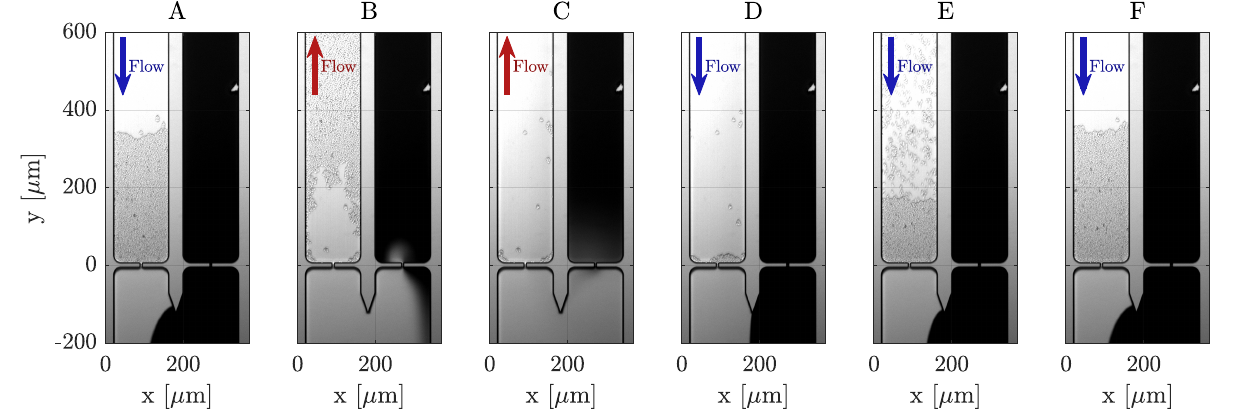}
  \caption{Presentation of the backflush experiments. One backflush starts at time $t_{B_i}$ (image A), at the moment when the flux is rapidly reverted, by setting the pressure $\Delta P_1$ at $-6.9$ kPa (image B) for 20 s (C). The pressure is finally reset to its initial value of $100$ kPa, and the clog rapidly returns to its initial position (D-F). Experience realized with $\Delta P_1 = 100$ kPa and $\Delta P_2 =$ 300 kPa.}
  \label{res:BKF-Presentation}
\end{figure*}

The response time of the present flow measurement technique can be approximated using the advection time necessary to change the interface shape when a variation of flow rate $Q_1$ is detected. For $Q_1\approx100\,$nL.min$^{-1}$ and an interface's region of interest of about 300\,$\mu$m, we estimate the response time between 0.1 and 0.2\,s, which is lower than the response time of commercial flow rate sensors. The main limitations of this method are (i) the need for specific calibration and (ii) a design-specific method, as the measurement microchannel must be implemented during chip microfabrication.

There is no apparent limitation on the minimal measured flow rate, except the pressure controller's ability to generate such very low flow rates. Our study's main experimental challenge was to avoid air bubbles in the microfluidic tubing. As the flow rate of our device is very low, it is possible to observe contact line pinning, which degrades the pressure control of the system from 10 Pa to several hundreds of Pa. To avoid this problem, a dedicated experimental protocol was set up. It was based on the direct connection of the reservoirs to the chips (without any valve, as it was noted that closing/opening them released tiny air bubbles), starting with a dry system and pushing the liquids slowly inside the device. This method could be easily implemented in other microfluidic studies, especially the ones involving filtration phenomena.

\subsection{Experimental protocol}
In this subsection, we briefly present the two different kinds of experiments carried out in this article (constant pressure experiments and backflush experiments) before turning to the results in Section III.

\subsubsection{Experiments at constant pressure}
During constant pressure experiments, the pressure $\Delta P_1$ is fixed at a constant value, while the clog and the interface are imaged at the same time, as illustrated in Fig. \ref{res:ConstructionIllustration}\,(top row). The flow rate ratio is measured as presented in section \ref{section:OnChipFlowRate}. 

The flow rate through the clog, its hydrodynamic resistance, and its length are computed for every picture recorded. The values associated with the example experiment presented in Fig. \ref{res:ConstructionIllustration}\,(top) are presented in Fig. \ref{res:ConstructionIllustration}\,(bottom). The clog length ($L_c$) increases with time and reaches $380\, \mu$m $20$ min after the beginning of clogging. More precisely, the speed of growth of the clog (i.e. the derivative of the clog length) decreases with time, as the clog reaches 100 $\mu$m in less than 3 min but needs more than 14 min to reach 300 $\mu$m. The flow rate ($Q_1$) rapidly decreases from $474$ nL.min$^{-1}$ at the beginning of the experiment, where no clog is present in the device, to less than 100 nL.min$^{-1}$ after 8 s. It continues to decrease with time, but more slowly, and reaches around 12.3 nL.min$^{-1}$ at the end of the experiment. In the same way, the hydrodynamic resistance of the clog ($R_c$) rapidly increases and exceeds the hydrodynamic resistance of the empty chip after only 2 seconds (picture A) and reaches a final value of $24.7 \times 10^{16}$ Pa.s.m$^{-3}$.

\subsubsection{Backflush cycles}
The second kind of experiment consists in backflush cycles. For 3 min, the pressure $\Delta P_1$ is kept fixed at a given value (generally at $100$ kPa) before being rapidly reverted at $-6.9$ kPa for 20 seconds, as illustrated in Fig. \ref{res:BKF-Presentation}. {\color{referee1} Under these conditions,} when the flow is reverted, the clog is abruptly destroyed, and, as the channel is $\approx 7$ mm long, the backflushed yeasts remain inside the device. After 20 seconds, the pressure returns to its initial value, and a new clog is  constructed again, {\color{referee1} in a few minutes (as it is much longer to build a clog than to destroy it)}.

Theoretically, the clog length and hydrodynamic resistance could be measured throughout the process. However, the suspended cell density can be very high during the re-construction phase (see Fig. \ref{res:BKF-Presentation}, panel E). Therefore, the values of hydrodynamic resistance obtained during this process are impacted by both (i) the hydrodynamic resistance of the clog and (ii) the hydrodynamic resistance of the dense yeast suspension being pushed in a confined channel. Therefore, in this article, the values of clog length and hydrodynamic resistance will be presented only at the \emph{final} state of the backflushes cycles when the clog is fully stabilized.

%%%%%%%%%%%%%%%%%%%%%%%%%%%%
%%%%%%%%%%%% RESULTS %%%%%%%%%%%%%%%
%%%%%%%%%%%%%%%%%%%%%%%%%%%%
\section{Experimental results}
\label{section:Results}
This section is divided into two parts. First, the relationship between the hydrodynamic resistance and the length of the clog is detailed under different pressure conditions. Next, the number of cells per unit volume is measured, and the results are detailed for different experimental conditions.

\subsection{Relationship between hydrodynamic resistance and clog length}
\label{subsec:Rh=fL}

\begin{figure*}
\centering
  \includegraphics[width=0.45\linewidth]{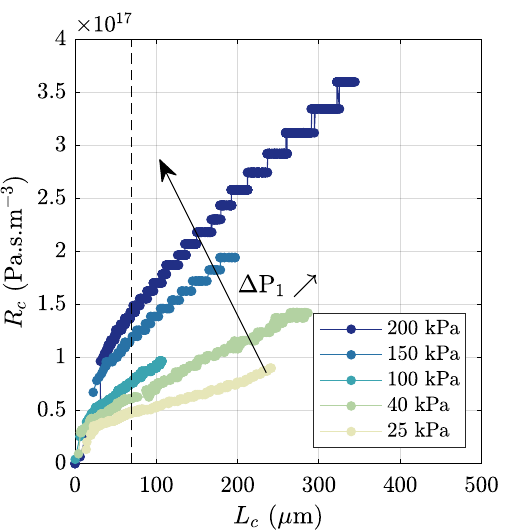}
    \includegraphics[width=0.45\linewidth]{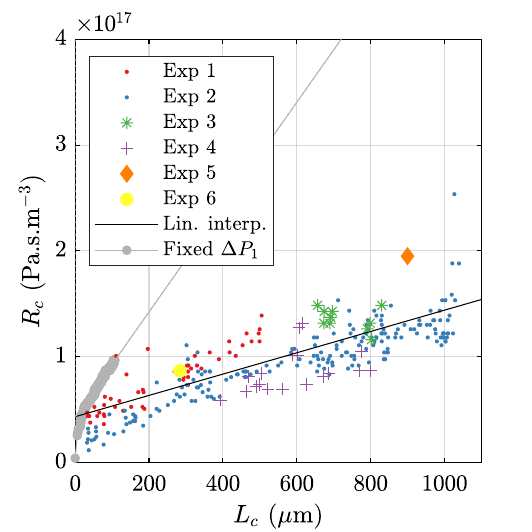}
  \caption{Hydrodynamic resistance of different clogs ($R_c$) constructed under different conditions, computed as a function of clog length ($L_c$). Results obtained during constant pressure experiments (left) and backflush cycles (right). For constant pressure experiments, different operating pressures are used, see the legend. A vertical dashed line is placed at $70\,\mu$m. Only a representative fraction of all the experiments are presented, as explained in the text. The operating pressure is kept at 100 kPa during {the reconstruction phase of the clog for backflush experiments}. The hydrodynamic resistance of backflushed clogs and the clog length are measured at the end of the backflush cycle, and a solid line represents a linear fit of all experimental data. With backflushes, experiments 4 and 6 were conducted in 6.35 $\mu$m-deep chips; the others were conducted in 6.1$\mu$m -deep chips. The data from the constant pressure experiment conducted at 100 kPa are reported in gray for easier comparison {(with extrapolated data from a linear fit  presented with a gray straight line)}.}
  \label{res:Rh_fL}
\end{figure*}

\begin{figure}
\centering
  \includegraphics[width=\linewidth]{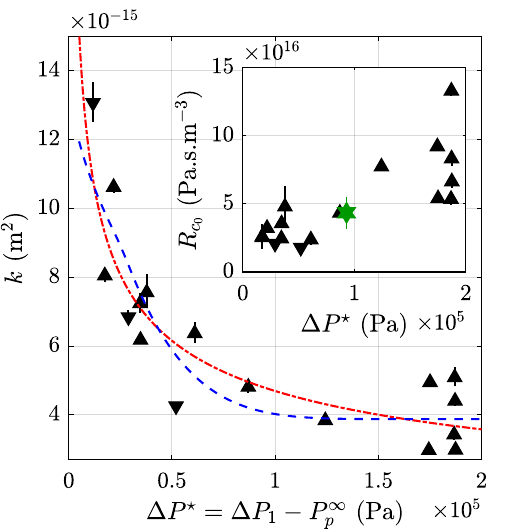} \\
  \includegraphics[width=\linewidth]{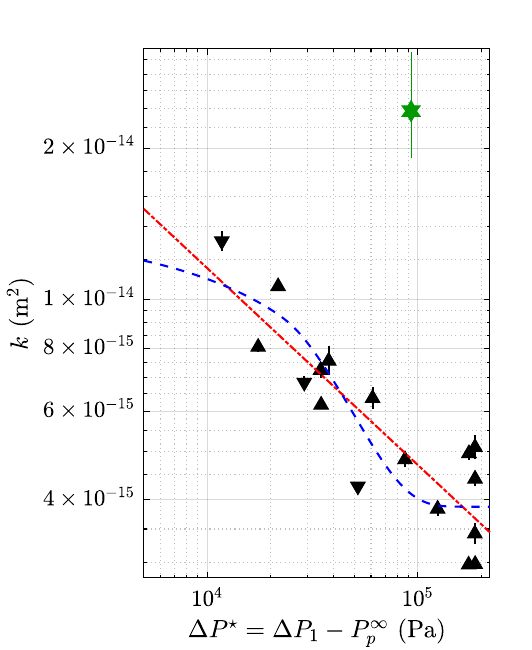}
  \caption{Permeability at constant pressure (triangles pointing upwards for experiments on a $6.1\,  \mu$m deep chip and downwards for experiments on a $6.35\, \mu$m chip), and during backflush cycles (green hexagram), with uncertainty bars representing the confidence interval at 99$\%$, computed during the linear fit. Linear representation at the top and log-log representation at the bottom. Dashed-dotted red line: fit from power law, equation \ref{eqn:powerlaw}. Dashed blue line: fit from the model developed in this article (equation \ref{eqn:Permeability3Termes}). In the top panel, the figure presents the dataset obtained in the range $[0.3 \times 10^{-14} - 1.5 \times 10^{-14}]$ m$^2$, excluding the permeability after backflushes. {\color{referee2} The inset presents the evolution of the clog offset hydrodynamic resistance ($R_{c_0}$) as a function of $\Delta P^\star$.}}
  \label{res:Permeability}
\end{figure}

As the solid volume fraction of the yeast suspension is not precisely the same from one experiment to another (cell density varies in the range $[5\times10^5-10^6]$ cells.mL$^{-1}$, see section \ref{sec:MatEtMet}), {\color{referee1} the evolution of the hydrodynamic resistance is presented with regards to the clog length rather than time}, as presented in Fig. \ref{res:Rh_fL}. During the initial phase of the clog construction, while it does not reach the inlet channel lateral walls ($L_c < 70\, \mu$m), the hydrodynamic resistance increases rapidly, from zero to more than $5\times10^{16}$ Pa.s.m$^{-3}$. As the clog length increases ($L_c > 70\, \mu$m), the hydrodynamic resistance increases linearly with length, as is expected if the clog behaves like a porous medium of constant permeability. The experiments are repeated at different $\Delta P_1$. All the clogs behave qualitatively as described before. However, the slope of the linear trend depends on $\Delta P_1$, especially from $25$ kPa to $100$ kPa. 

In the same way, we realized 284 backflush cycles during six independent experiments. Each independent experiment is conducted with a different yeast suspension, and in a different chip. For a given experiment, the final length (and hydrodynamic resistance) is not always constant between 2 cycles. Indeed, the consecutive backflush cycles are non-symmetrical: more flow goes from the inlet to the outlet than vice versa. Some yeasts may accumulate on the top of a backflushed clog between two successive backflush cycles. {\color{referee1} As a consequence, the clog length changes between two consecutive cycles. Therefore, for a given experiment, several points are obtained. Some of the experiments (numbers 3 to 6) have been conducted with an already-formed clog.}

The results are presented in Fig. \ref{res:Rh_fL}, (right) where the clog hydrodynamic resistance is shown as a function of its length. As with constant-pressure experiments, the hydrodynamic resistance is found to increase with clog length, while can be noticed that the resistance values are much smaller than during constant-pressure experiments: after backflushes,  for $L_c=  1000 \, \mathrm{\mu m}$, the hydrodynamic resistance is around $1.5 \times 10^{17}$ Pa.s.m$^{-3}$, while this value is reached at constant-pressure for $L_c \approx 300\,\mu$m, with $\Delta P_1 = 40$ kPa (see Fig. \ref{res:Rh_fL}).

\subsection{Permeability}
\label{subsec:Permeability}
The Darcy formalism is adopted to characterize the hydrodynamic resistance of a clog. The permeability is then computed from the slope of the hydrodynamic resistance following the equation:
\begin{equation}
R_c = R_{c_0} + \frac{\eta L_c}{k S_t},
\end{equation}
\noindent with $R_{c_0}$, a hydrodynamic resistance offset, that we interpret as a consequence of effects that occur between 0 and $70\, \mu$m, $\eta$ the dynamic viscosity of the solution, $L_c$ the clog length, $S_t$ the transverse surface of our device (here $\approx 6\, \mu\mathrm{m} \times 140\, \mu$m) and $k$ the clog permeability. The data obtained before $70\, \mu$m are {not taken into account for the linear fit}.

Fig. \ref{res:Permeability} presents {\color{referee1}all} the values of permeability obtained during constant pressure experiments as a function of the difference of hydrodynamic pressure experienced by the clog ($\Delta P^\star$), together with the permeability value extracted from all the backflush experiments conducted with $\Delta P_1=100$ kPa. \textcolor{referee2}{Fig. 9 (top) and Fig. 9 (bottom) are linear and logarithmic representations respectively.} \textcolor{referee2}{The inset of Fig. 9 (top) presents the values of the clog offset hydrodynamic resistance $R_{c_0}$ obtained during constant pressure experiments as a function of $\Delta P^\star$, together with the $R_{c_0}$ value extracted from all the backflush experiments conducted with $\Delta P_1=100$ kPa.} 

When the clog is constructed at constant pressure, the permeability is maximum {under} $25$ kPa, reaching more than $10 \times 10^{-15}\, \mathrm{m}^{2}$, and is severely decreased as operating pressure reaches more than $100$ kPa, where the permeability is lower than $6 \times 10^{-15}\, \mathrm{m}^{2}$. No significant change in permeability is observed between $100$ kPa and $200$ kPa, {\color{referee1} where it ranges from $2.9\times 10^{-15}$ m$^2$ to $5.37\times 10^{-15}$ m$^2$}. \textcolor{referee2}{The clog offset hydrodynamic resistance $R_{c_0}$ has an increasing trend from $2\times10^{16}$ to $13\times10^{16}\,$Pa.s.m$^{-3}$ with relatively large experimental dispersion.}

The permeability of clogs constructed during backflush cycles reaches $2.37 \times 10^{-14} \text{ m}^2$, with a 99\% confidence interval of $[1.91 \times 10^{-14}, 3.10 \times 10^{-14}]$ m$^2$, see green hexagram data point in Fig. 9. These values are about four times higher than the permeability of a clog constructed by yeast accumulation at a constant pressure of $100$ kPa and higher than all the previously measured permeability values. \textcolor{referee2}{The value of $R_{c_0}$ for clogs constructed during backflush cycles reaches $4.4\times10^{16}$\,Pa.s.m$^{-3}$, which is similar to the constant-pressure experiments.}

\subsection{Cell density measurements}
\label{section:MICRO}
\begin{figure}
\centering
    \includegraphics[width=\linewidth]{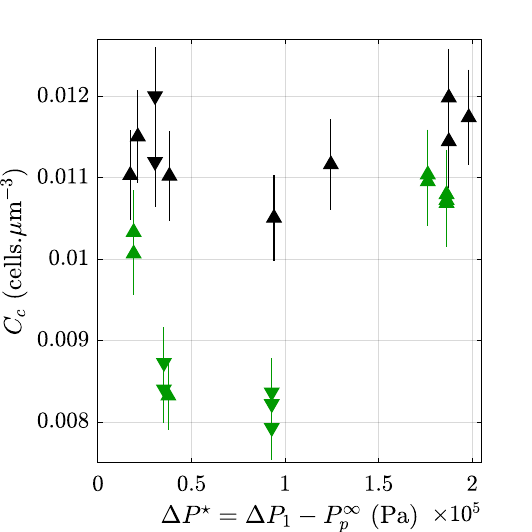}
  \caption{Cell density as a function of $\DPS$, for the two different kinds of experiments: constant pressure in black and backflush cycles green. Experiments conducted on a 6.1 $\mu$m chip are noted with upwards pointing triangles and those conducted on a 6.35 $\mu$m chip are noted with downwards pointing triangles. Error bars represent errors of 5\% of the mean value.}
  \label{res:MICRO-C=f(DP)}
\end{figure}

The cell density is computed for clogs obtained during constant-pressure experiments or after backflush, for different $\Delta P^\star$. The results are presented in Fig. \ref{res:MICRO-C_f(DP)}. During constant pressure filtration, the cell density is between 10.5 $\times 10^{-3}$ and 12.0 $\times 10^{-3}$ cell.$\mu$m$^{-3}$, and no apparent impact of $\DPS$ can be observed. After backflushes, the cell density is lower than before backflushes. This time, the operating pressure seems to have an impact, but this impact is non-monotonic, as the cell density is around $10.3 \times 10^{-3}$ cell.$\mu$m$^{-3}$ at 25 kPa, 8.3$\times 10^{-3}$ cell.$\mu$m$^{-3}$ at 100 kPa and $10.7 \times 10^{-3}$ cell.$\mu$m$^{-3}$ at 200 kPa.

\section{Modeling permeability in constant pressure experiments}
\label{sec:Model}

We propose now to model the constant-pressure experiments presented above. First, we apply the classical scaling used in previous bioclogging studies performed at the membrane scale. Then we develop a model based on physical considerations that challenges the classical (bio)filtration theory.

\subsection{Classical (bio)filtration theory}
The dependence of permeability with operating pressure is a widely reported phenomenon in the scientific literature related to filtration. The permeability is usually not computed in these studies, but the specific hydrodynamic resistance is preferred for practical reasons. This variable is defined as follow: 
\begin{equation}
\alpha = \dfrac{R_c S_t^2}{m \eta},
\end{equation}

\noindent where $m$ is the total mass of the deposited yeasts, $S_t$ the transverse surface of the device, and $\eta$ the liquid dynamic viscosity. The use of this metric is due to the experimental protocol of these studies, where the filtration cake is hardly observable, but the deposited mass on the filter can easily be weighted. With this variable, the effect of pressure on $\alpha$ is often expressed following an empirical power law \citep{rushton_solidliquid_1996,foley_review_2006}: 
\begin{equation}
\alpha = \alpha_0 (\DPS)^n,
  \label{eqn:powerlaw}
\end{equation}
\noindent where $\alpha_0$ and $n$ are fitting parameters. $\alpha_0$ represents the specific resistance at 1 Pa, whose value is rarely discussed, and n is the ``compressibility index'', whose value is of interest. When n=0, the cake is incompressible ($\alpha$ does not depend on $\DPS$), and when $n>0$, the cake is said to be compressible. Reported values \citep{meireles_origin_2004,valencia_direct_2022,rushton_filtration_1977,nakanishi_specific_1987,tanaka_factors_1993} of $n$, associated with dead end filtration of \textit{Saccharomyces cerevisiae} range from 0.25 to 1.10. As the cell density is not significantly affected by the operating pressure (see section \ref{section:MICRO}), the mass of a given clog of a given length should not be impacted either. The following relationship allows the conversion of $k$ to $\alpha$:
\begin{equation}
k=\frac{V}{\alpha m},    
\label{eqn_keqfalpha}
\end{equation}
where V is the volume occupied by the clog (defined by the transverse section of the device $S_t$ multiplied by the clog length $L_c$). It follows that equation \ref{eqn:powerlaw} can be re-written as:
\begin{equation}
    k = k_0 (\DPS)^{-n},
    \label{eqn:powerlaw_k}
\end{equation}
with $k_0$ the permeability at 1 Pa.

Finally, after fitting the experimental points obtained in this study, we compute a compressibility index of 0.39, with confidence bounds at 95\% of $[0.28 - 0.50]$ (see the fitted curve in Fig. \ref{res:Permeability}), which is in the literature range. The quality of the fit can be quantified with the coefficient of determination: $R^2=0.816$ and the root mean square error $RMSE=1.134 \times 10^{-15}\, \mathrm{m}^2$. One may note that the confidence interval at 95\% is relatively large. We believe this is because the compressibility index is not constant over our entire dataset. If the points between 0 and 75 kPa are used (and the others are ignored), the fitted index equals 0.52. When only the points obtained between 75 and 200 kPa are used, the index equals 0.23. The fact that the compressibility index may itself depend on $\DPS$ is largely reported in the literature of microbial filtration   \citep{foley_review_2006,nakanishi_specific_1987,katagiri_yeast_2021}. Some authors recommend the use of the equation \ref{eqn:powerlaw} only above a certain pressure \citep{katagiri_yeast_2021}, but without discussing the choice of this pressure, taken as an \textit{ad hoc} value.

\subsection{Another model, based on physical considerations}
\subsubsection{Model presentation}
The increase of resistance with pressure is widely accepted to be a consequence of interstitial space contraction. However, there are no physical reasons or scientific consensus about the quantitative relationship between the hydrodynamic resistance and the operating pressure \citep{foley_review_2006}.

More generally, two kinds of modeling are reported in the literature to relate the microstructure of a particle packing to the permeability. The most simple one is the Kozeny-Carman equation \citep{carman_fluid_1937}.
To establish this equation, the authors first assume that the pore network throughout the packing can be approximated by a series of microchannels of a given radius and length. 
An analytical relationship is then derived, in the case of rigid and monodisperse packing of spheres, to express the permeability as a function of porosity ($\phi$) and particle diameter ($d$):
\begin{equation}
k = \dfrac{\phi ^3 d^2}{36 K_k (1-\phi)^2},
\label{EQN:KC}
\end{equation}
\noindent with $K_k$ an empirical coefficient. This equation has been widely used to model the permeability of yeast clogs, even if there is no consensus on the correct value of $K_k$. Some authors \citep{meireles_filtration_2002,meireles_origin_2004} have modified this equation to take into account the fact that the permeability can tend toward zero while the porosity is still higher than zero. 
The second kind of modeling is pore network modeling \citep{xiong_review_2016, chareyre_pore-scale_2012}. In that case, the interstitial space is modeled as a collection of inter-connected pores, and the full topology of the pore network is simulated. Each channel is characterized by a given hydrodynamic conductivity \footnote{Reminder: the permeability is a large-scale quantity that can be computed as soon as the porous medium can be seen as a homogeneous medium. The permeability of a porous medium made of several parallel channels is proportional to the sum of the hydrodynamic conductivity of each of these parallel channels.} and a given length and is connected to its neighbors. In the case of polydisperse spheres \citep{oren_process_2002,catalano_pore-scale_2014,chareyre_pore-scale_2012}, this kind of modeling gives very satisfactory results, as long as the definition of the effective pore throat radius ($r_e$) and the pore connectivity is properly estimated. In that case, the individual hydrodynamic conductivity of each pore is proportional to $r_e^4$.

\begin{figure}[t]
    \centering
    \includegraphics[width=\linewidth]{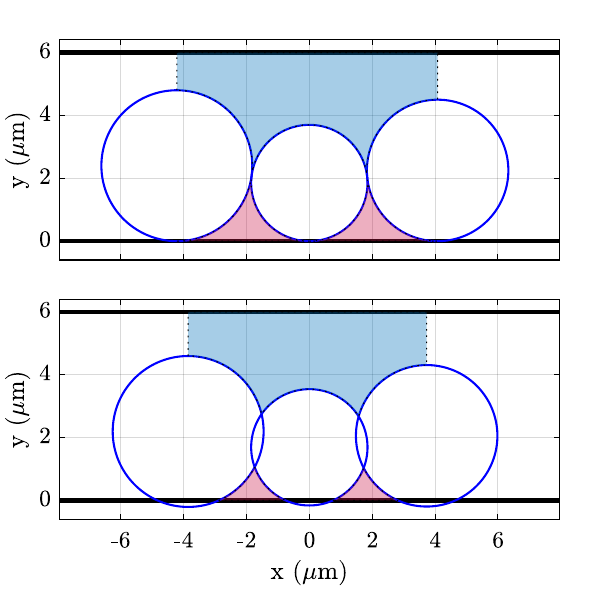}
    \caption{Representation of the two different kinds of pore openings in the yeast clog: small pores, located between pairs of cells and the wall (red), and large defects  (blue).}
    \label{fig:FigureSchemaPetitsGrandsPores}
\end{figure}

In this article, we describe an intermediate model using arguments from both the Kozeny-Carman relationship and the pore network modeling. The experimental system of this article is highly confined (the height of the device is around 6 $\mu$m, slightly larger than the diameter of a particle). Therefore, the walls of the device create two important effects, {illustrated in Fig. \ref{fig:FigureSchemaPetitsGrandsPores}}. First, since the particles are in contact with the channel's walls, the clog presents bigger pores than they would have inside a non-confined packing. In particular, we assume that most of the pores are located between a pair of particles and the confining wall. Therefore, we will only analyze the contribution of the cells in contact with the walls and neglect the contribution of the pores located away from the top/bottom walls.  Besides, we will also neglect the influence of the lateral walls - as they are located 140 $\mu$m away from each other. We also assume that the effective radius $r_e$ is equal to the radius of the pore throat located between each pair of cells and the wall ($r_t$). Second, the confinement produces geometrical exclusion: two cells of diameter larger than $3\,\mu$m cannot be aligned in the vertical direction, as the depth of the device is only equal to $\approx 6\, \mu$m. This geometrical exclusion creates large voids in the packing. The analytical resolution of the typical length scale of these voids is far from trivial (one needs to consider the wide variety of random 3D arrangements that such packing may create), and we do not attempt to model them here. However, one can estimate that their typical length scale is large compared to the individual deformation of one yeast cell, so it should not depend on $\DPS$. For example, in the case where one cell of $5\,\mu$m of diameter is in contact with a wall, on the top of it, there is a void of $\approx 1\,\mu$m in height and of several $\mu$m in width. These large voids constitute a pore network whose permeability is not affected by the clog compression induced by a pressure increase. To summarize, it is assumed that there are two independent systems of pores: one whose dimensions can be affected by the compression of the clog, and one with large pores, insensitive to compression. The permeability is therefore modeled by the equation:
\begin{equation}
k = A\times r_t^4 + B,
\label{eqn:Permeability3Termes}
\end{equation}
\noindent where $A$ and $B$ are parameters, fitted against all the experimental points available (Fig. \ref{res:Permeability}). Physically, the term $A\times r_t^4$ represents the compressible system of pores, and $B$ represents the large pores, unaffected by mechanical compression. The precise relationship between $r_t$ and $\DPS$ is given in Supplementary Information 2. The main assumptions needed to analytically derive this relationship are:
\begin{itemize}
    \item {Yeast cells are approximated by spheres that deform locally, with no deformation away from the contact zone,} so that their deformation is represented by intersecting spheres.
    \item Yeast cells are polydisperse, and their size distribution is based on experimental data.
    \item The effective pore radius is assumed to be equal to the radius of the pore throat.
    \item {At low deformation, yeast cells behave as Hertzian balls \citep{chang_evaluating_2021}, where the deformation is proportional to the applied force at the power $\frac{2}{3}$. At high deformation, cells behave as pressurized elastic shells, as in Vella et al. (2012) \citep{vella_indentation_2012}, where the deformation is proportional to the applied force.}
    \item The contact forces are assumed to be homogeneous in the packing (in other words, the contact forces are the same for all the particles subjected to a given compressive stress).
    \item {\color{revisions2}The compressive stress is assumed to follow the predictions of the linear poroelasticity theory {(see \cite{biot_general_1941}, \citep{wang_theory_2001}} or \citep{macminn_large_2016} for example), where the compressive stress is linearly distributed in the packing}.
\end{itemize}

\subsubsection{Comparison with experimental data}

{The coefficients obtained when fitting our experimental data against equation \eqref{eqn:Permeability3Termes} are $A=1.12 \times 10^{11}\, \mathrm{m}^{-2}$  with a  $95 \%$ confidence interval of $[8.11 \times 10^{10} - 1.43 \times 10^{11}]\, m^{-2}$ and  $B=3.88\times 10^{-15}\, \mathrm{m}^{2}$ with a  confidence interval of $[3.00\times 10^{-15} - 4.76 \times 10^{-15}]\, m^2$. With these values, $R^2=0.80$ and $RMSE = 1.19\times 10^{-16} \mathrm{m}^2$. The quality of this fit is similar to the one obtained with the power law (see curves in Fig. \ref{res:Permeability}).
When only the data points between 0 and 75 kPa are considered, the fitted parameters are $A = 1.29\times 10^{11}\, \mathrm{m}^{-2}$ and $B = 3.12\times 10^{-15}\, \mathrm{m}^{2}$, close to the previously determined parameters. When only the points between 75 kPa and 200 kPa are used, then the value of $A$ is badly determined – which is related to the fact that for this pressure range the compressible pores are completely closed (this will be further discussed in the next section), and $B = 3.94 \times 10^{-15}\, \mathrm{m}^{2}$, again very close to the previous values.}

{A rough value of A can be estimated as $N \frac{2\pi}{S_t \tau}$ {(see Chareyre et al. \citep{chareyre_pore-scale_2012})}, with $\tau$ the tortuosity of the flow channels connecting the pores present between neighboring cells and the microchannel top/bottom walls and $N$ the total number of such channels. Taking $\tau = \frac{\pi}{2}$  (the tortuosity value for a porous medium consisting of monodisperse spheres) and $N = 32.5$ (that is the  140 $\mu$m channel width divided by the mean yeast cell diameter 4.3 $\mu$m), one gets $A = 1.55\times 10^{11}$ m$^{-2}$ which is relatively close to the fitted value.  
The value of B can also be estimated using some scaling arguments, although in a less satisfactory manner. The hydraulic resistance of a rectangular flow channel that would connect the large defects in the packing is roughly equal to $R^r = \frac{12 \mu L \tau}{h^3 w}$, with $L$ the length of the channel, $h$ its height, and $w$ its width. Because of the term $h^3$, a wide range of values of $B$ can be reproduced by small changes of $h$. Therefore, the physical significance of the $w$ and $h$ values is questionable.  But we note that the fitted value of B can be reproduced, for instance, by considering 28 straight channels with $ h= 0.8\, \mu$m and $w=3\, \mu$m, or 6 channels of width dimensions $h=1.1 \, \mu$m  and  $w=4 \,\mu$m. All these numbers are realistic values.}

%%%%%%%%%%%%%%%%%%%%%%%%%%%%%
%%%%% Discussion %%%%%%%%%%%%
%%%%%%%%%%%%%%%%%%%%%%%%%%%%%
\section{Discussion}
\label{sec:Discussion}
\subsection{Permeability and hydrodynamic resistance of a yeast clog constructed at constant pressure}

When the yeast clog is constructed under constant pressure, the permeability of a yeast cake is found to be strongly impacted by the operating pressure (Fig. \ref{res:Permeability}), a result already largely reported at the macroscale. Interestingly, no change in cell density could be observed (see Fig. \ref{res:MICRO-C=f(DP)}). It is likely due to slight differences in the microstructure that we are unable to measure. For example, the closing of a pore throat significantly decreases the permeability but is very difficult to detect in the density signal because of the yeast polydispersity. 

The permeability values measured in the present study {correspond to specific resistances (see equation \ref{eqn_keqfalpha})} in the lower range of what is reported in the literature. Between 100 kPa and 200 kPa, we observe a specific resistance between $2\times 10^{11}$ m.kg$^{-1}$ and $3\times 10^{11}$ m.kg$^{-1}$, where Nakanishi et al. (1987) observed a specific resistance of $\approx 5.2\times 10^{11}$\,m.kg$^{-1}$ at 140 kPa \citep{nakanishi_specific_1987}, and Mota et al. \citep{mota_modeling_2012} observed a specific resistance of $\approx 4.2\times 10^{11}$ m.kg$^{-1}$, while other studies report higher values (see  Meireles et al. \citep{meireles_origin_2004}). {This can be explained by the fact that, in the filtration literature, studies are usually carried out in a non-confined geometry so that the wall effects are negligible (a point clearly stated in the theoretical development of Tien and Romaro (2013) \cite{tien_can_2013}). In that case, the parameter B decreases, which means the overall permeability decreases, and consequently, the specific resistance increases.
} The compressibility index ($n$) obtained by adjusting our data with a  power law is also in the range of what is observed in literature for dead-end filtration of \textit{Saccharomyces cerevisiae} \citep{meireles_origin_2004,valencia_direct_2022,rushton_filtration_1977,nakanishi_specific_1987,tanaka_factors_1993} but it depends on the range of data used for the fit (see subsection IV.A).

\begin{figure}
     \centering
     \includegraphics[width=\linewidth]{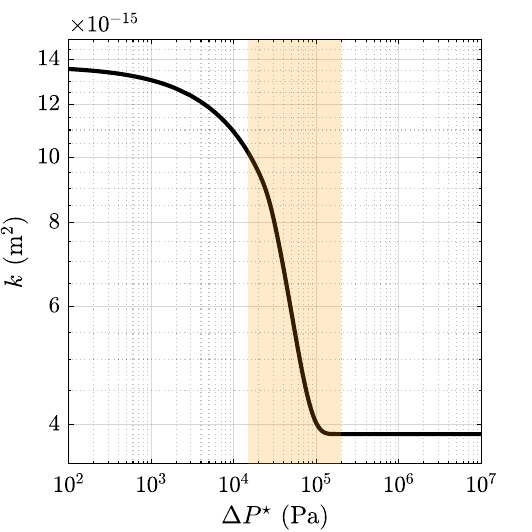}
     \caption{Permeability predicted by the fitted model, presented in section \ref{sec:Model}. The operating pressure varies in the range [10$^2$ - 10$^7$] Pa, and the permeability is simulated over the full range of operating pressure. The shaded area represents the pressure range explored during experiments.}
     \label{fig:DiscussionModel}
 \end{figure}

To further explore this point, the model presented in subsection IV.B.1 is used to simulate the permeability of a yeast clog over an extended range of pressures (i.e. from 10$^2$ Pa to 10$^7$ Pa). The results are presented in Fig. \ref{fig:DiscussionModel}.  Between 0 and approximately 20 kPa, our model predicts that the deformations of the cells are negligible (permeability is constant). Between 20 and 100 kPa, the model predicts that the pore throats close significantly, while after 100 kPa, the pores' throats are completely closed. At this point, the permeability equals B, the residual permeability due to the poorly compressible large defects in the packing, which is independent of $\DPS$.

This behavior of the model is consistent with two observations made in the literature, to which we have already alluded.  
First, over a limited pressure range, the results from the model can be adjusted locally by a power law (i.e. a straight line on the log-log graph shown in Fig. 12), with a compressibility index (slope of the straight line) that clearly depends on the data range used for the fit. 
Second, the fact that some authors report that the classical power law is applicable only after a certain pressure \citep{nakanishi_specific_1987,katagiri_yeast_2021}, reflects the fact that the pressure needs to reach a certain value before the deformation of cells becomes significant. This is provoked by a combination of (i) the geometry of the pore space and (ii) the mechanical behavior of the yeast cells. In turn, that mechanical behavior is controlled by the osmotic pressure at the moment of filtration, which implies taking care of this point when designing the experimental protocol, as it was done in the present study.

%\textcolor{referee1}{Whereas we note a large change in permeability in the studied pressure range, we did not observe a clear change in clogs' density between constant-pressure experiments. The porosity of such a clog can be written as $\phi=1-V_{1c}C_c$ where $V_{1c}$ is the volume of one cell (about 65\,fL according to Alric \emph{et al.} \cite{alric_macromolecular_2022}). If we consider a simple Kozeny-Carman model of the permeability-porosity relation, an increase of 10\% of the cell density $C_c$ (which is in the error bars of our measurements, Fig. \ref{res:MICRO-C=f(DP)}) leads to a clog's permeability divided by 3.}

%%%Paul : qques ajouts /rephrasage ci-dessous > OK
\textcolor{referee1}{Whereas we note a large change in permeability in the studied pressure range, we did not observe a concomitant significant  change of the cell density $C_c$ within clogs, see black data points in Fig. 10. The porosity of such  clogs can be written as $\phi=1-V_{1c}C_c$ where $V_{1c}$ is the volume of one cell (about 65\,fL according to Alric \emph{et al.} \cite{alric_macromolecular_2022}). If we consider a simple Kozeny-Carman model of the permeability-porosity relation, an increase of 10\% of the cell density $C_c$ (which is in the error bars of our measurements, see Fig. \ref{res:MICRO-C=f(DP)}) leads to a reduction of the clog permeability by a factor 3.  Therefore, it's possible for the permeability of the clogs to change in an important way, with only a small change in cell density.}

%All of these elements may explain why authors report that the classical power law is applicable only after a certain pressure \citep{nakanishi_specific_1987,katagiri_yeast_2021}. There is a limit where the deformation becomes significant, which is a combination of (i) the geometry of the pore space and (ii) the mechanical behavior of the yeast cells. In turn, that mechanical behavior is controlled by the osmotic pressure at the moment of filtration, which implies taking care of this point when designing the experimental protocol, as it was done in the present study.

Finally, for a clog length between 0 and $70\,\mu$m, the hydrodynamic resistance increases rapidly and non-linearly, see Fig. \ref{res:Rh_fL} (left panel). As a consequence, the linear fit reveals an offset: $R_c = \frac{\mu L_c}{k S_t}+R_{c_0}${\color{referee2}, with an increasing trend of $R_{c_0}$ in the experimentally considered pressure range.} Several hypotheses could explain {\color{referee2} the presence of} this offset. First, it has been reported that the interaction of the first particles and the membrane pore can have a strong impact on the final hydrodynamic resistance of a yeast clog \citep{meireles_origin_2004}. Second, it has been reported that the presence of a porous medium can alter the curvature of the flow paths in the channel so that the apparent permeability of a thin deposit layer is increased \citep{dufreche_apparent_2002}.
{\color{referee2} Also, the observed trend could be a consequence of cell deformation at the pore vicinity. Finally, the dispersion} of $R_{c_0}$ could be a consequence of the polydisperse nature of the yeast cells, interplaying with these processes. A better characterization of the mechanisms that control the value of $R_{c_0}$ is beyond the scope of the present study.
%%%Paul : je ne comprends pas l'ajout de  {\color{referee2}The observed trend could be a consequence of cell deformation at the pore vicinity
% est-ce abordé dans le retour au referee ? ici cela tombe un peu comme un cheveu au milieu de la soupe...a minima commencer la phrase par un "Also,..." > OK. On a une réf dans la réponse aux referee à la modif de la discussion, ça me semble bien comme ça. J'ai rajouté Also.

\subsection{Impact of construction history - backflush cycles}
The construction history has a very important impact on the permeability of the clog. The results of
this article show that the permeability is greatly enhanced by imposing backflush cycles. To give
statistical significance to this result, many experiments and backflush cycles were realized at a fixed
$\DPS$ ($\approx$ 100 kPa), see Fig. 8, right panel. However, to assess the impact of the
construction pressure on the clog microstructure and therefore on the permeability, some cell
density measurements were also carried out after backflushes conducted at different $\DPS$, see
green data points in Fig. 10.

The cell density after a backflush is found to depend on $\DPS$ in a non-monotonic manner, with the
cell density being the lowest at intermediate pressure ($\approx 100$ kPa). Therefore, in this latter
case, the increase in permeability is likely related to this decrease in cell density. This is consistent
with the physical interpretation we proposed in section \ref{sec:Model}. A smaller cell density
suggests that the packing is looser, involving larger size for the pores located in between the particles
and walls, and for the larger defects as well. The decrease in cell density may originate from the fact
that the clog is made of a loose packing of cell aggregates rather than individual cells, these
aggregates being formed during the backflush. This idea is consistent with what is reported in
the literature, as it is documented that pre-aggregating yeast cells, with the help of a flocculation
molecule, can significantly reduce the hydrodynamic resistance during dead-end filtration
\citep{chandler_high_2004} or cross-flow filtration \citep{wickramasinghe_enhanced_2002}.
The presence of aggregates cannot be assessed with certainty on the images recorded after backflush
(see Fig. \ref{res:BKF-Presentation}, panel B or E) as it is impossible to distinguish aggregated cells
and cells only in close contact on such images. However, a literature review supports the hypothesis
of aggregate formation/fragmentation, see Introduction.

To explain why the cell density after backflush for high and low pressure is close to the one measured
in the constant-pressure experiments, we propose the following interpretation. The yeast-yeast adhesion
mechanism has been described at the protein scale \citep{el-kirat-chatel_forces_2015}, and it is
reported to work like a Velcro. It requires no energy to stick two adhesive proteins, but much energy
is necessary to unbind two adherent proteins. Besides, the adhesion forces are proportional to the
area of the contact surface, as each cell is surrounded by a relatively uniform layer of these adhesive
proteins. Therefore, the adhesion force scales as $\sqrt{\delta}$, with $\delta$ the deformation of a
cell. These adhesive forces compete against repulsive forces (described precisely in section
\ref{sec:Model}). These forces scale as $\delta^{3/2}$ at low deformation and as $\delta$ at high
deformation. At low pressure ($< 40 $ kPa), two cells are put in contact together with a very low
force, so that the number of proteins that have been bound is low: it is easy to separate the cells
from each other. This is going to limit the formation of lasting aggregates during the backflush.
At very high pressure
($\approx 200$ kPa), when cells are pushed very firmly against each other, the repulsion forces
between pair of cells become larger than the adhesion forces (as $\delta^{3/2} \gg \delta^{1/2}$ and
$\delta\gg\delta^{1/2}$ when $\delta \rightarrow \infty$). This will weaken the aggregates which
may fragment during the clog re-construction stage, at high pressure. Finally, cell aggregates are
present with a significant impact on the clog density only in the intermediate pressure range.

%%%%%%%%%%%%%%%%%%%%%%%%%%%%%%%%
%%%%% CONCLUSIONS %%%%%%%%%%%%%%
%%%%%%%%%%%%%%%%%%%%%%%%%%%%%%%%

\section{Conclusions and perspectives}
\label{sec:Conclusion}
In this article, we presented a study of bioclogging by yeast cells in a microfluidic chip. Experimental developments allowed us to measure the hydrodynamic resistance and thus the permeability of a confined yeast clog in a precise and robust way. Our method is based on flow comparison and image recognition – where the images of a colored interface are compared to a precalibrated image database - and this method can measure flow rates under 10 nL.min$^{-1}$ with good accuracy and response time. The clog permeability, deduced from the clog’s hydrodynamic resistance and length, first decreases sharply as the operating pressure increases, before reaching a plateau. The experimental data points can be fitted by a power law, as it is usually done in the literature without any physical basis, but the exponent of the power law is very sensitive to the data range used, in terms of operating pressure. A model based on physical arguments from the physics of granular media, fluid mechanics, yeast cell mechanical behavior, and geometry is presented and gives similar results as the power law in terms of adjustment to the experimental data. It assumes the co-existence of two-pore networks in the yeast clog. The first one represents the pores between neighboring cells and the microchannel top/bottom walls. Radii of these pores are very sensitive to the operating pressure. The second one connects the large defects in the yeast cell packing resulting from the polydispersity of the yeast suspension and the strong confinement. Radii of these pores are much less affected by the operating pressure: its contribution to the clog permeability is supposed to be constant.  Our results also emphasized that the ``history’’ of the clog construction is a crucial parameter, as clogs constructed following a backflush have a much larger permeability. We believe this is because, in this case, the clog is made of a loose packing of cell aggregates rather than individual cells, and this is supported by the measurement of the cell density in the clog. 

The work presented in this article opens many perspectives. From a modeling point of view,  achieving a description of the whole pore network existing in the present 3D and highly confined yeast cell packing could support the picture of the two-pore networks above, on which our interpretation of the data is based.  Also, we assumed that the permeability of the clog was constant once its extent in the flow direction was large enough, typically larger than $70 \, \mathrm{\mu m}$. However,  the permeability might be spatially non-uniform as the layers of cells near the membrane are expected to experience higher mechanical constraints \citep{macminn_large_2016,meireles_origin_2004}, and this may offer some directions to refine the present modeling. \textcolor{referee1}{Future work taking into account the 3D assembly of cells and the poromechanical coupling inside the clog could also investigate the relationship between cell deformation, cell density and clog permeability.}

%The work presented in this article opens many perspectives. From a modeling point of view,  achieving a description of the whole pore network existing in the present 3D and highly confined yeast cell packing could support the picture of the two-pore networks above, on which our interpretation of the data is based.  Also, we assumed that the permeability of the clog was constant once its extent in the flow direction was large enough, typically larger than $70 \, \mathrm{\mu m}$. However,  the permeability might be spatially non-uniform as the layers of cells near the membrane are expected to experience higher mechanical constraints \citep{macminn_large_2016,meireles_origin_2004}, and this may offer some directions to refine the present modeling.  

From an experimental point of view, one may tune the microfluidic experimental design to vary the confinement, e.g. to check that less confined clogs experience larger permeability variations as is expected from the present modeling. Indeed, the contribution to the permeability of the pore network based on the large packing defects close to the top/bottom wall is expected to decrease when the confinement is relaxed. More details about the clog microstructure could be obtained using cells with fluorescent cytoplasm or an exclusion fluorescence technique \citep{delarue_self-driven_2016,bottier_dynamic_2011}. Also, a more systematic study of the clog reconstruction after a backflush is needed, focusing on the impact of the backflush characteristics (pressure, duration) on the clog hydrodynamic resistance in relation to the properties of the resuspended clusters of cells.

Our results were obtained using a microfluidic filtration cell with a unique constriction, whose size is larger than the average particle size. It is the first time such a geometry is used to study the filtration of a biological suspension. Indeed, pores are usually much smaller than the cell size, because filtration aims to actually \emph{retain} biological particles and because these particles may deform to a certain extent and pass through some constrictions that would be not small enough. The generalization of our result to a more realistic multipore membrane is not straightforward as one can expect a strong influence of the membrane geometry (pore size and density) from situations where a single yeast cell completely blocks an isolated pore \citep{meireles_origin_2004} to situations where a given cell deposited close to a given pore impairs the clogging of neighboring pores by steric exclusion (a mechanism known as ``pore protection phenomenon'' \citep{valencia_direct_2020}). Depending on the situation, the effect of the first layer of deposited cells on the total hydrodynamic resistance of the clog may be more or less predominant. 
 
Finally, it should be remembered that all these results were obtained with non-proliferating cells and may be extended to situations where the effects of proliferation are negligible (for example, if filtration spans shorter time scales than the proliferation time scale). However, generalizing these results for proliferating cells might be challenging, as cell proliferation is complex, affecting cell properties and interacting with the other processes involved in clogging.  Indeed, to proliferate, cells need nutrients that are diluted in the liquid phase flowing through the clog. This flow depends on the clog microstructure, which in turn may be significantly altered by proliferation, as it has been reported that proliferation provides energy to mix the clog locally \citep{ranft_fluidization_2010}.  This coupling and its impact on the clog permeability is important to decipher as some filtration processes involving living, proliferating cells can extend over timescales comparable to or longer than the proliferation timescales.

%%%REFERENCES%%%
\bibliographystyle{apsrev4-2}
\bibliography{Biblio_soumission}

%%%%%%%%%%%%%%%%%%%% Supplementary %%%%%%%%%%%%%%%%%%%%
\newpage
\setcounter{figure}{0}
\renewcommand{\figurename}{Fig. SI}

\section*{Supplementary Information 1: Sensitivity analysis of the method}
In this Supplementary Information, we detail some sensitivity tests about our on-chip flow rate measurement method, before precisely quantifying uncertainty magnitudes for the different measurements that are presented in this article.

\subsection{Sensitivity analysis}

\begin{figure}
\centering
  \includegraphics[width=\linewidth]{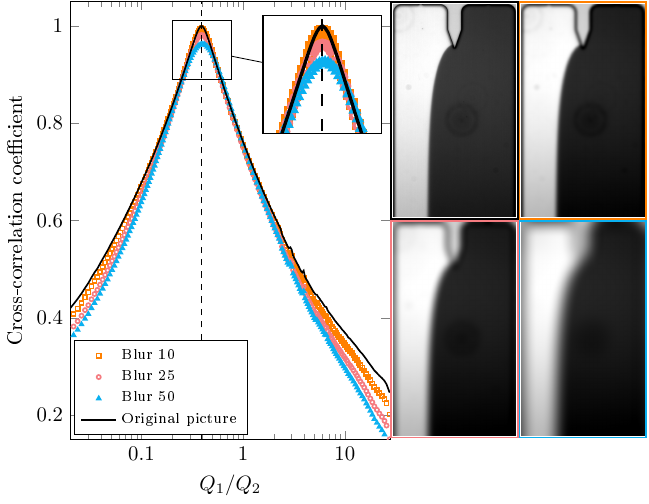}
  \caption{Influence of a blur. The dashed line represents the expected $Q_1/Q_2$ ratio.}
  \label{si:flou} 
\end{figure}

Fig. SI \ref{si:flou} shows that adding an artificial blur to the image has almost no effect on the cross-correlation vs flow rate ratio graph. Blur is added using a Gaussian filter with a standard deviation of 10, 25, and 50 pixels.

\begin{figure}[h!]
\centering
  \includegraphics[width=\linewidth]{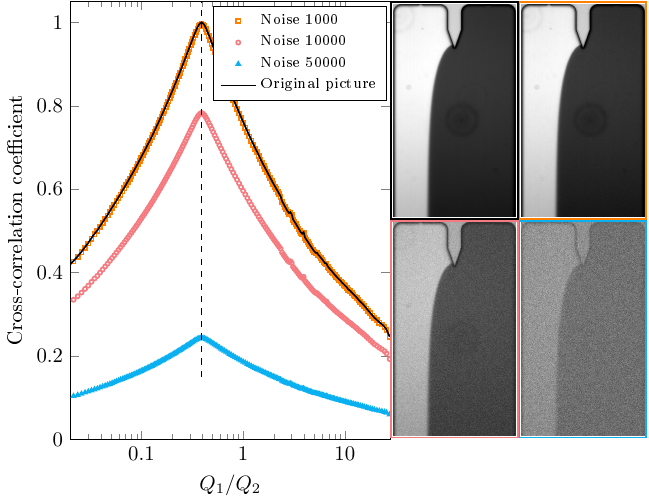}
  \caption{Influence of white noise. The dashed line represents the expected $Q_1/Q_2$ ratio.}
  \label{si:noise}
\end{figure}

Fig. SI \ref{si:noise} shows that adding artificial white noise to the image has no effect on the flow rate detection as the peak remains sharp. Nevertheless, the cross-coefficient value of this peak decreases when adding noise. White noise of amplitude 1000, 5000, and 10000 is added to 16-bit encoded pixels intensity.
 
\begin{figure}[h!]
\centering
  \includegraphics[width=\linewidth]{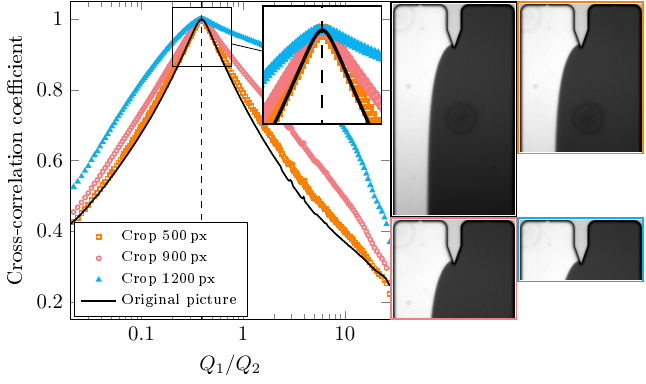}
  \caption{Influence of cropping the image, for a flow rate ratio close to 1. The dashed line represents the expected $Q_1/Q_2$ ratio.}
  \label{si:crop_center}
\end{figure}

Fig. SI \ref{si:crop_center} shows that cropping the bottom part of the picture, when the colored fluid-non-colored fluid interface is close to the center of the microchannel, has no effect on the flow rate detection as the peak remains sharp and its value is still close to 1. We observe some deviations for the most important crop but far from the peak. The bottom part of the picture is cropped of 500\,px, 900\,px, and 1200\,px.

\begin{figure}[h!]
\centering
  \includegraphics[width=\linewidth]{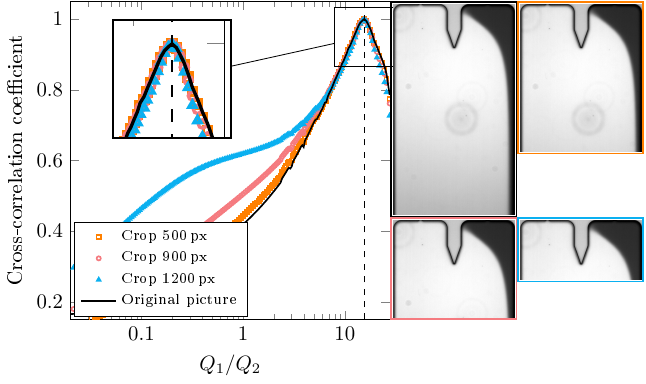}
  \caption{Influence of cropping the image, for a flow rate ratio far from 1. The dashed line represents the expected $Q_1/Q_2$ ratio.}
  \label{si:crop_decale}
\end{figure}

Fig. SI \ref{si:crop_decale} shows that cropping the bottom part of the picture, when the colored fluid-non-colored fluid interface is close to the right-hand wall of the microchannel, does not affect the flow rate detection as the peak remains sharp and its value is still close to 1. We observe some deviations for the most important crop but far from the peak. The bottom part of the picture is cropped of 500\,px, 900\,px, and 1200\,px.

These different sensitivity tests show the robustness of our image recognition method.

\subsection{Uncertainties}

In terms of uncertainty, an error in the geometrical characteristics of the microchannels can lead to an error in the hydrodynamic resistance. The uncertainty on H is estimated to be about 1\%. A quick analysis of uncertainty propagation shows that a  1\% uncertainty in $H$ leads to a 3\% uncertainty in the hydrodynamic resistance of the device. This uncertainty affects $R_i$ and $R_p$ in the same way  (i.e. an overestimation of $H$ will lead to an underestimation of \emph{both} $R_i$ and $R_p$). So the resulting uncertainty on $\gamma$, and therefore $Q_1/Q_2$, is very low. For instance, between $H = 6.1\,\mu$m and $H = 6.35\,\mu$m, there is a difference of $4 \%$, but the difference on the values of $\gamma$ is $0.31 \%$. In the same way, the uncertainty associated with $P_p^\infty$ is very low, below $0.5\%$. 

Uncertainties on applied pressure are $10\,$Pa according to the data sheet of the pressure controller. For $\Delta P_1$ and $\Delta P_2$ around 50\,kPa, the induced uncertainty on $Q_1/Q_2$ is negligible (0.04\%).

Besides, the discretization induced by a finite-size data bank gives an uncertainty on the actual flow rate ratio, as it could be between two pictures in the data bank. The flow rate ratio separation between two successive images gives an estimate of the uncertainty. We estimate that it corresponds to an uncertainty between $3\%$ and $6 \%$ on $Q_1/Q_2$, depending on the flow rate ratio (with the highest uncertainty at low/high $Q_1/Q_2$). This could easily be reduced by increasing the number of pictures in the database, at a higher computational cost.

{
That database discretization is thus directly responsible for an uncertainty up to $6\%$ on the clog hydrodynamic resistance.} But this uncertainty is not systematic, as the clog resistance increases with time, so the effects of discretization on the uncertainty on the permeability $k$ are much lower than $6 \%$. This can be observed in Fig. \ref{res:Rh_fL}. In the curve obtained at $200$ kPa, the values are \emph{quantified} to certain levels, which is a consequence of the database discretization. To quantify the uncertainty associated with this phenomenon, we compute the confidence interval at 99\% during the linear fit. The associated values are represented as error bars in Fig. \ref{res:Permeability}. Note that, in addition to these error bars, values of permeability reported in this article are precise at $\pm 3\%$, because of a potential error on the height of the device, that creates a systematic bias of $3\%$ on the value of $R_c$.

Regarding the cell detection algorithm, used to compute the yeast cell density ($C_c$), the uncertainty is estimated by comparing two images of the same clog: one taken slightly out of focus and the other perfectly focused. In this case, the algorithm gave two different values of density separated by 5\%. Therefore, for the sake of robustness, we estimate that the precision of the cell density is 5\%.

Finally, a precise summary of these quantities is listed in table \ref{tbl:Uncertainties}, for each variable presented in this article. When two uncertainty sources are independent (for example $H$ and the database discretization), the following formula is used to compute the final magnitude: 
\begin{equation}
    U_m = \sqrt{U_1^2+U_2^2}
\end{equation}
{with $U_m$ the``final'' uncertainty magnitude, $U_1$ and $U_2$ the uncertainty magnitudes associated with independent variables. For example, the uncertainty of $R_c$ is equal to $\sqrt{6^2+3^2} \approx 6.7 \%$. These final uncertainty magnitudes are also listed in table \ref{tbl:Uncertainties}.
}
\section*{Supplementary Information 2: Permeability modeling}

\begin{figure}[t]
    \centering
    \includegraphics{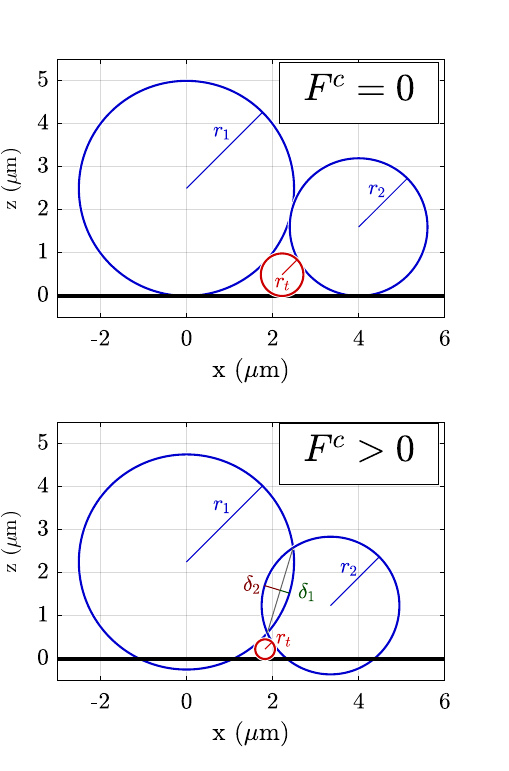}
    \caption{Geometrical representation of the pore throat, for two cells of different radius $r_1$ and $r_2$.}
    \label{si:D_Modelgeom}
\end{figure}

The modeling of the variation of $k$ as a function of $\DPS$ is done in four steps. 
Firstly, the effective radius is expressed as a function of the pore throat radius ($r_t$) in a particular configuration, representative of our system, so that a relationship between $k$ and $r_t$ is obtained. Secondly, the pore throat radius is expressed as a function of cell deformation ($\delta$). Thirdly, the variation of cell deformation with contact forces ($F^c$) is derived. Fourthly, the magnitude of contact forces is linked to the operating pressure ($\Delta P^\star$). Altogether, this workflow allows us to relate  the permeability of a confined packing and the operating pressure and  can be summarized as follow:
\[    k \longleftrightarrow r_t \longleftrightarrow \delta \longleftrightarrow F^c \longleftrightarrow \DPS 
\]
\noindent with the symbol $\longleftrightarrow$ representing functions that will be detailed successively in the following. The final expression $k \longleftrightarrow \DPS$ is cumbersome and provides no interesting information for this article, but a Matlab script that computes the relationship is provided as supplementary material.
\paragraph{$\bm{r_t \longleftrightarrow \delta}$}
Yeast are ovoid objects with a low aspect ratio ($\approx 1.1$), so their geometry can be approximated by spheres. As the particles are in contact with the walls, the general equation governing the radius of the pore throat is derived as follows. If two spherical particles of radius $r_1$ and $r_2$ are in contact with each other and with a wall, then, in the plane of contact, their position can be fully determined by two pairs of coordinates: $(x_1,y_1)$ and $(x_2,y_2)$, respectively (see illustration on Fig. SI \ref{si:D_Modelgeom}). The largest sphere that can be inserted between these two spheres and the wall has coordinates $(x_t,y_t)$ and radius $r_t$ that are solutions of the following system of equations:
\begin{equation}
    \begin{cases}
    (x_1-x_t)^2 + (y_1-y_t)^2 = (r_1+r_t)^2 \\
    (x_2-x_t)^2 + (y_2-y_t)^2 = (r_2+r_t)^2 \\
    \end{cases},
    \label{EQN:SystemRt}
\end{equation}
\noindent where the wall is assumed to be a line of equation $y=0$.

In the literature, it is reported that when two yeasts are pushed slowly against each other, they deform like a pressurized elastic shell \citep{vella_indentation_2012}. At relatively low applied forces, it has been observed that yeasts behave like Hertzian balls \citep{chang_evaluating_2021} so that the deformation is negligible away from the contact zone. At higher forces, it has been reported that the volume of yeasts decreases when they are compressed, as water flows \emph{out} of the cytoplasm \citep{smith_uniquely_1998, vella_indentation_2012}. In that case, Vella et al. \citep{vella_indentation_2012} showed that the deformation of a yeast cell is negligible far away from the indentation zone. Therefore, we represent the deformation of neighboring yeasts as overlapping spheres that do not change their shape/radius as they are compressed: when two cells of radius $r_1$ and $r_2$ are compressed against each other and against the walls, they may overlap their neighbor by an amount $\delta_1$ and $\delta_2$, respectively (see Fig. SI \ref{si:D_Modelgeom}, bottom panel for an example). In that case, $y_1 = r_1-\delta_1$, $y_2 = r_2 - \delta_2$, and $y_t = r_t$. Without loss of generality, one can fix $x_1 = 0$, so that $x_2$ is fully determined from $r_1$, $r_2$, $\delta_1$ and $\delta_2$ :
\begin{equation}
    x_2 = \sqrt{ (r_1+r_2-\delta_1 - \delta_2)^2 - (r_1-r_2)^2 }.
\end{equation}
With these elements, the system (\ref{EQN:SystemRt}) can be analytically solved, so that $r_t$ is known for every possible configuration ($r_1$,\,$r_2$,\,$\delta_1$,\,$\delta_2$):
\begin{equation}
        r_t = \dfrac{1}{(2 r_1- \delta_1)}\left(\dfrac{ x_2-\sqrt{ \lambda }  }{    (2-2\dfrac{2 r_2-\delta_2}{2 r_1-\delta_1})} \right)^2 - \dfrac{\delta_1}{2},
        \label{EQN:rt}
\end{equation}
with 
\begin{equation}
     \lambda = x_2^2 - 4 (1-\dfrac{2 r_2-\delta_2}{2 r_1-\delta_1}) ((r_1-\delta_1)(r_2-\delta_2)+(2 r_2-\delta_2)(\delta_1-\delta_2)).
     \label{EQN:uglyrt}
\end{equation}

\begin{figure}
\centering
  \includegraphics[width=\linewidth]{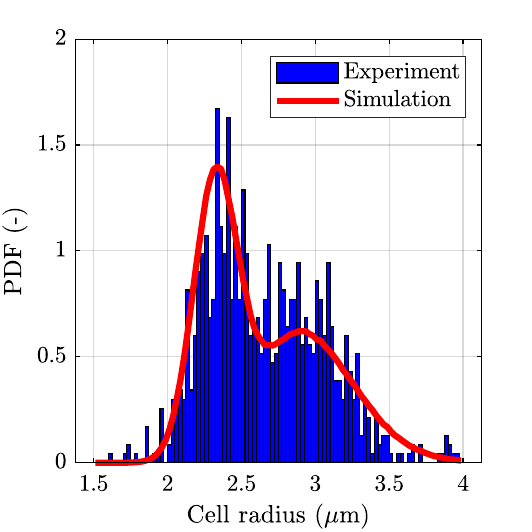}
  \caption{Size distribution of yeast cells: experimental data (blue bars) and modeling of the size distribution as a sum of two normal distributions (red line).}
  \label{si:polydisp}
\end{figure}

{\color{revisions2}Yeast cells are polydisperse, so the combination of $r_1$ and $r_2$ is random. Experimentally, a simple experiment is conducted to estimate polydispersity. A drop of yeast suspension is put between a microscope glass slide and its coverslip, and the yeast cells are observed under a microscope. The cell size distribution has been measured thanks to a simple segmentation algorithm, where the area of each cell is computed. The equivalent radius is then given by $r = \sqrt{\mathcal{A}/\pi}$, with $\mathcal{A}$ the segmented area. The results are presented in Fig. SI \ref{si:polydisp}. They show that the size distribution of yeast cells is indeed polydisperse and presents two local maxima: one around 2.33 $\mu$m, the other around 2.93 $\mu$m. This allowed us to conclude that, in our experiment, half of the yeast cells were budding \footnote{Yeast cells reproduce by budding. That is to say that a daughter cell grows from the surface of a mother cell. The radius of the mother cell remains (almost) constant during that process. In contrast, the radius of the daughter cell increases rapidly until a point where the daughter cell separates from the mother cell.}. The size distribution of mother cells and non-budding cells is assumed to follow a bimodal distribution, well described by two Gaussian distributions: the first one represents cells with no buds, while the other one represents cells with one bud. The distribution of cells without buds is first fitted and described by a normal law with a mean of 2.35 $\mu$m and a standard deviation of 0.15 $\mu$m. At the same time, the other part of the bimodal distribution represents the pairs of particles, whose area is equal to $\mathcal{A} = \pi(r_m^2+r_b^2)$, with $r_m$ and $r_b$ the radius of a mother cell and a bud, respectively. From that observation, the size distribution of daughter cells is simulated as a normal law of mean equal to 1.8 $\mu$m and of the standard deviation of 0.5 $\mu$m. Finally, these two normal laws reproduce well the observed size distribution - if one assumes that there are as many cells without buds as there are cells with buds. In other words, one-third of the cells are actually daughter cells.  

Finally, to take into account the cell size distribution, every result is averaged over $10^4$ simulated combinations of ($r_1, r_2$). More precisely, the permeability is equal to
\begin{equation}
    k = A\times \langle r_t(\delta_1,\delta_2)^4 \rangle_{(r_1,r_2)} + B 
\end{equation}
\noindent with $\langle r_t(\delta_1,\delta_2)^4 \rangle_{(r_1,r_2)}$ the mean of the throat radius given by formula \ref{EQN:rt}, computed over all the combinations of $(r_1,r_2)$. With no deformation ($\delta_1 = \delta_2 = 0$), this mean is equal to $8.864 \times 10^{-2}\, \mu$m$^{4}$, which corresponds to what would be obtained with a (monodisperse) pore throat radius of 0.55 $\mu$m. This radius ``at rest'' is relatively close to what would be obtained without considering the polydispersity. Indeed, in the monodisperse case, the pore throat radius is equal to:
\begin{equation}
    r_t = \dfrac{r_1^2 + 2\delta_1^2 - 4 r_1 \delta_1}{4 r_1 - 2 \delta_1}
\end{equation}
{Thus, at rest, $r_t = 0.25 r_1$. Therefore, $\langle r_t ^4 \rangle = \langle (0.25 r_1)^4 \rangle =  8.864 \times 10^{-2}\, \mu$m$^{4}$, which gives an equivalent pore throat radius of 0.54 $\mu$m.}

\paragraph{$\bm{\delta \longleftrightarrow F^c}$} 
The relationship between the deformation of a cell and the compressive force is extracted directly from the literature. As already mentioned earlier, at low forces, Chang et al. \citep{chang_evaluating_2021} provide experimental data on the flat indentation of yeast cells. From their paper, it can be derived that the deformation is given by:
\begin{equation}
    \delta(F^c | F^c \leq F^c_{lim}) = \left( \dfrac{ 3 F^c (1-\nu^2)}{2 E \sqrt{r}} \right)^{2/3} 
    \label{EQN:Delta_LF}
\end{equation}
with  E the Young modulus of the cell at low deformation (that they estimated to be around 4.5 MPa), and $\nu$ the Poisson coefficient (that they estimated to be equal to 0.5), $F^c_{lim}$ the limit between low forces and high forces, and $r_i$ the radius of the cell. At large deformations, Vella et al. \citep{vella_indentation_2012} provide analytical relationships for the indentation of a yeast cell with a punctual geometry. From their paper, we derive that, at large deformations: 
\begin{equation}
    \delta(F^c | F^c \geq  F^c_{lim}) = \dfrac{F^c}{2\pi P_t r} + \delta_0
    \label{EQN:Delta_HF}
\end{equation}
{\color{revisions2}with $P_t$ the turgor pressure of a yeast cell (that is estimated around 0.1\,MPa, based on published results \citep{vella_indentation_2012}, \citep{alric_macromolecular_2022}), }and  $\delta_0$ an unknown constant. The $\delta_0$ constant is estimated from equation \ref{EQN:Delta_LF} and \ref{EQN:Delta_HF}, assuming that the force-deformation law is continuous: 
\begin{equation}
\delta_0 = \left( \dfrac{ 3 F^c_{lim} (1-\nu^2)}{2 E \sqrt{r}} \right)^{2/3}  -  \dfrac{F^c_{lim}}{2\pi P_t r} 
\end{equation}
Following Vella et al., (2012) \citep{vella_indentation_2012}, the limit between small deformations and large deformations is defined at $\approx 100\,\mathrm{nm} = \delta_{lim}$. This corresponds to a limit between small forces and high forces equal to:
\begin{equation}
F_{lim}^c = \dfrac{2 E\sqrt{r}}{3 (1-\nu^2)}\delta_{lim}^{3/2}
\end{equation}

\begin{figure}
\centering
  \includegraphics[width=\linewidth]{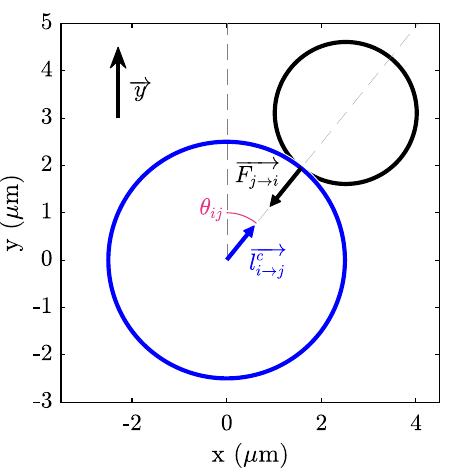}
  \caption{Schematic representation of the branch vector and the contact force between a pair of particles $i$ (in blue) and $j$ (in black).}
  \label{si:branch}
\end{figure}

\paragraph{$\bm{F^c \longleftrightarrow \DPS}$} The relationship between the contact forces exerted on each pair of particles and the mean stress is known to follow the relationship \citep{latzel_macroscopic_2000}:
\begin{equation}
    \langle \boldsymbol{\sigma} \rangle = \dfrac{1}{V}\sum_{i \in V} w_i^{V} \sum_{j=1}^{Z_{i}} \overrightarrow{ F_{j \rightarrow i}^c} \otimes \overrightarrow{l^c_{i \rightarrow j}}
\end{equation}
\noindent with $\otimes$ the diyadic product, $\langle \boldsymbol{\sigma} \rangle$ the mean stress tensor computed over the volume $V$, $w_i^{V}$  the weight corresponding to the fraction of the volume of particle $i$ which lies inside the averaging volume $V$, $Z_i$ the number of contacts for particle $i$, $\overrightarrow{ F_{j \rightarrow i}^c}$ the contact force between particles $j$ and $i$ (see example presented Fig. SI \ref{si:branch}) and  $\overrightarrow{l^c_{i \rightarrow j}}$ the ``branch vector'', which is defined from the center of particle $i$ with radius $r_i$ to the contact point between particle $j$ and $i$, see Fig. SI \ref{si:branch}. It is often assumed that $\overrightarrow{l^c_{i \rightarrow j}} \approx r_i \overrightarrow{n^c_{{i \rightarrow j}}}$, with $\overrightarrow{n^c_{{i \rightarrow j}}}$ the contact normal vector from particle $i$ to particle $j$ \citep{latzel_macroscopic_2000}. In our case, the packing is compressed by uniaxial compression (say $\overrightarrow{y}$ direction) carried out by a fluid. Therefore, the stress component of interest is $<\sigma_{yy}>$, and, in the absence of friction, $\overrightarrow{ F_{j \rightarrow i}}^c$ is colinear to $\overrightarrow{l^c_{i \rightarrow j}}$, and both vectors have the same angle ($\theta_{ij}$) with respect to the $\vec{y}$ direction. Therefore, the former equation reduces to  :
\begin{equation}
    \langle \sigma_{yy} \rangle = \dfrac{1}{V}\sum_{i \in V} w_i^V \sum_{j=1}^{Z_{i}} F_{j \rightarrow i}^c \,  r_i \, cos^2(\theta_{ij})
\end{equation}

The outcomes of this modeling effort demonstrated a favorable level of quantitative agreement when compared to empirical observations. Consequently, in the present work, we have chosen to draw inspiration from their approach and make the assumption that forces are uniformly distributed throughout the packing.

The forces in disordered granular media are known to be heterogeneous. But previous work \citep{louf_under_2021}, investigating the swelling of hydrogel beads in a packing of glass beads, has developed a model to better understand this swelling, simplifying the stress field following a mean-field approximation. This modeling has shown good quantitive results when compared to experimental data. Consequently, we have decided to adopt their approach and assume that forces are evenly distributed throughout the packing (with our notations, this implies that $F_{j \rightarrow i}^c \approx F^c$, with $F^c$ the mean contact force). By defining the volume properly so that $w_i^V=1$ for all particles considered in the volume $V$, the former equation then becomes :
\begin{equation}
    \langle \sigma_{yy} \rangle = \dfrac{1}{V}\sum_{i \in V} F^c  r_i \sum_{j=1}^{Z_{i}} \cos^2(\theta_{ij})
    \label{EQN:Equation15}
\end{equation}

The last part of the sum is then approximated by:
\begin{equation}
    \sum_{j=1}^{Z_{i}} \cos^2(\theta_{ij}) \approx Z_i \langle cos^2(\theta_{ij}) \rangle_{\theta_{ij} \in [0-2\pi]} = \dfrac{Z_i}{2}
\end{equation}
So that equation \ref{EQN:Equation15} becomes:
\begin{equation}
     \langle \sigma_{yy} \rangle \approx \dfrac{F^c}{2V}\sum_{i \in V}  r_i Z_i
\end{equation}
$r_i$ and $Z_i$ are then assumed to be two independent random variables, so that $ \langle r_i Z_i \rangle_{i \in V}$ = $ \langle r_i \rangle _{i \in V} \langle Z_i \rangle _{i \in V}$, where $ \langle r_i \rangle _{i\in V}$ and $ \langle Z_i \rangle_{i \in V}$ are the mean radius and coordination number, respectively (noted simply $\bar{r}$ and $\bar{Z}$ after). With these assumptions, the final relationship between the force exerted between each pair of particles and the ambient compressive stress is :
\begin{equation}
     F^c \approx \dfrac{2  \langle \sigma_{yy} \rangle }{\bar{r}\bar{Z}} \dfrac{V}{n_{i \in V}} = \dfrac{2 \langle\sigma_{yy}\rangle}{\bar{r}\bar{Z} C_c}
\end{equation}
where $C_c$ is the number of cells per unit volume (estimated from measurements presented in Fig. \ref{res:MICRO-Illustration}).

{\color{revisions2}Finally, the stress field distribution inside the clog is not known at this point. Therefore, we follow one of the main assumptions of linear poroelasticity: the stress field is assumed to be linearly distributed through the porous medium, with $ \sigma_{yy}  = 0$ at the top of the clog and $\sigma_{yy}=\DPS$ close to the membrane so that $\sigma_{yy} \approx \DPS (1- y/L)$ (see for example MacMinn et al. \citep{macminn_large_2016}).  Therefore $\langle \sigma_{yy} \rangle \approx \DPS /2 $ and the relationship between $F^c$ and $\DPS$ is}:
\begin{equation}
     F^c \approx  \dfrac{\DPS}{\bar{r}\bar{Z} C_c}
\end{equation}

{\color{revisions2}This closes the system of equations and provides an analytical relationship between $k$ and $\DPS$. However, because of equations \eqref{EQN:rt} and \eqref{EQN:uglyrt}, that relationship is cumbersome to write. The final (numerical) relationship between $k$ and $\DPS$ is shown in Fig. \ref{fig:DiscussionModel}, where the values of the parameters $A$ and $B$  are those obtained by adjustment to the experimental data.}

\end{document}